\documentclass[10pt,floatfix,
  aps,
  prx,
  amsmath,amssymb,
  superscriptaddress,
  nofootinbib,
  longbibliography,
  twocolumn
]{revtex4-2}

\usepackage{graphicx}
\usepackage{xcolor}
\usepackage{hyperref}
\usepackage{xurl}
\definecolor{linkblue}{RGB}{0,70,160}
\hypersetup{colorlinks=true, allcolors=linkblue}
\newcommand{\pillarhead}[2]{\medskip\noindent\textbf{#1.} \textit{#2}\par\smallskip\noindent}

\begin{document}

\title{Principles and Techniques for Agentic Research in Experimental Physics}

\author{Niko R. Reed}
\affiliation{Department of Physics, University of Maryland, College Park, Maryland 20742, USA}
\affiliation{Quantum Technology Center, University of Maryland, College Park, Maryland 20742, USA}
\affiliation{Joint Quantum Institute, University of Maryland, College Park, Maryland 20742, USA}

\author{Declan M. Daly}
\affiliation{Department of Physics, University of Maryland, College Park, Maryland 20742, USA}

\author{Andrew J. Beling}
\affiliation{Department of Physics, University of Maryland, College Park, Maryland 20742, USA}
\affiliation{Quantum Technology Center, University of Maryland, College Park, Maryland 20742, USA}

\author{Jordan D. Giandrea}
\affiliation{Department of Physics, University of Maryland, College Park, Maryland 20742, USA}

\author{Ronald L. Walsworth}
\affiliation{Department of Physics, University of Maryland, College Park, Maryland 20742, USA}
\affiliation{Quantum Technology Center, University of Maryland, College Park, Maryland 20742, USA}
\affiliation{Joint Quantum Institute, University of Maryland, College Park, Maryland 20742, USA}
\affiliation{Department of Electrical Engineering, University of Maryland, College Park, Maryland 20742, USA}

\author{John W. Blanchard}
\email{jwblanchard@umd.edu}
\affiliation{Quantum Technology Center, University of Maryland, College Park, Maryland 20742, USA}

\author{Daniel Ang}
\email{dga@umd.edu}
\affiliation{Quantum Technology Center, University of Maryland, College Park, Maryland 20742, USA}

\date{October 12, 2026}

\begin{abstract}
The use of artificial intelligence (AI) in physics research is increasingly common, yet few guidelines exist for its effective use. Starting with a principles-based approach, we outline a set of ten requirements for collaborative AI agents in experimental physics, spanning scientific norms, effective agent use, and domain expertise. We take a human-centered approach in which agents act as collaborators that amplify rather than replace the expertise of a human researcher (user) who directs the work and is responsible for its results. We then present a framework designed to enforce these requirements, detailing methods for improving off-the-shelf behavior of large language models (LLMs) with custom scaffolding and describing general-purpose scientific workflows and skills. In particular, we highlight the power of closed-loop campaigns where agents work iteratively toward a well-defined goal. We demonstrate the potential of this mode of research through six case studies in experimental physics, covering device design, hardware control, neural network optimization, physical model identification, and data analysis. All six case studies result in useful work products that add utility to our research, including a neural network architecture and solid-state quantum defect photodynamic model that substantially outperform our published work; a microwave waveguide design that extends a published design concept to a fabrication-ready device meeting our experimental constraints; a quantum control pulse sequence candidate that improves simulated performance by more than an order of magnitude; commissioning of an NMR spectrometer to $\sim$1~Hz frequency precision on live hardware; and finally a data-fitting pipeline that achieves, to our knowledge, the first reconstruction of the Ising microstate of an artificial spin-ice array from diffraction-limited magnetic imaging.
\end{abstract}

\maketitle

\section{Introduction}
\label{sec:intro}
Large language models (LLMs) underpin a growing set of powerful, general-purpose artificial intelligence (AI) tools for coding and problem-solving \cite{zhao_survey_2023}. Equipping models with structured instructions, tools, and file access enhances their capabilities and allows semi-autonomous work \cite{wang_survey_2024}; these arrangements are directly realized in popular harnesses such as Claude Code \cite{anthropic_claudecode} and Codex \cite{openai_codex}. Custom tools and workflows can enhance performance further still. Such LLM-based agents\footnote{Throughout, ``LLM'' denotes the underlying language model and ``agent'' an LLM equipped with tools, file access, and a persistent working loop.} have been employed in a wide variety of applications across physics and engineering, including symbolic reasoning and physical-model identification~\cite{nagele_agentic_2026}, analysis of real experimental data~\cite{moreno_ai_2026, wang_starwhisper_2025}, construction and automation of simulation workflows~\cite{zhang_agentic_2026, wang_humanenhanced_2026, lin_tomaster_2026}, design of devices and instruments~\cite{lupoiu_multiagentic_2025, berger_physicsintheloop_2026, du_turboagent_2026}, experimental protocol creation~\cite{arlt_autonomous_2025, panapitiya_autolabs_2025, hellert_agentic_2026}, and hands-on operation and tuning of real apparatus~\cite{xu_vibe_2026, cao_agents_2025, mandal_autonomous_2025, hu_electromagnetic_2025, sha_llmbased_2025,isogawa_agentic_2026}. A particularly powerful mode is the closed-loop agentic workflow, in which a model proposes candidate solutions, scores them with a frozen evaluator, reviews results, and plans the next iteration \cite{romeraparedes_mathematical_2023, novikov_alphaevolve_2025, aygun_ai_2026, zimmer_agentic_2026, karpathy_autoresearch_2026}. These loops enable sustained exploration of solution spaces far larger than a human researcher could search by hand, and have begun to be applied to physics research problems \cite{zhang_autoturb_2025, du_llm4ed_2024, grayeli_lasr_2024, xia_srscientist_2025, hu_multiagent_2026, yin_optimizing_2025,zoccheddu_large_2026,kaiser_large_2025, bosio_synthesizing_2025, vonhippel_refining_2025, sun_mephisto_2025, liu_repower_2025}.

However, fully realizing the promise of agentic research in experimental physics presents several challenges. First, models of physical systems often involve some degree of approximation whose correctness must be determined through empirical measurements and scientific judgment, unlike in fields such as pure mathematics and software development, where a formalized proof checker or test suite can quickly evaluate a proposed solution. Research problems in physics often involve complex sets of physical, experimental, and logistical constraints that require human judgment rather than the judgment of an agent acting alone. Experimental data add a unique challenge, as statistical noise, systematic effects and bias, and apparatus-specific information not available to the agent can cause errors and complicate the analysis. Second, even if a scientific result is correct, a useful research product requires careful tracking of the entire data collection and analysis pipeline, in contrast to other fields where a final product such as a proof or an algorithm may stand apart from the process that produced it. Appropriate verification and disclosure require well-documented code and methodology from all project phases, clearly delineated assumptions and sourced values, and an organized supporting file structure that a human can navigate and review. These key elements are not consistently preserved without specific guidance and structure to the agent. Finally, agentic research in physics is vulnerable to common failure modes of LLMs, including hallucinated numbers \cite{liu_autoresearch_2026}, metric gaming \cite{schwartz_vibe_2026, islam_gwbenchmarks_2026, ye_replicationbench_2025}, premature commitment \cite{nagele_agentic_2026, alfred_data_2026}, spatial reasoning errors \cite{wang_humanenhanced_2026, zhang_agentic_2026}, and promotion of unphysical candidate models \cite{yang_large_2025}.

As agentic research transitions from an object of study to a research tool, it is important to create replicable frameworks that address these issues and streamline useful work in experimental physics. Previous efforts in agentic research have demonstrated impressive proofs of concept~\cite{arlt_autonomous_2025,cao_agents_2025,hellert_agentic_2026}, but the prompts or frameworks employed are often highly specialized for a single class of problems or subfield, and offer few generalizable lessons. At the other extreme, some AI research has focused on autonomous research agents meant to handle the entire paper creation process from idea generation to manuscript preparation, with minimal human involvement \cite{lu_ai_2024,tang_airesearcher_2025,tang_fars_2026}, rather than on collaborative tools that support existing research efforts \cite{bisht_agentic_2026,nature_aiscientists_2026}; see Ref.~\cite{kong_ai_2026} for a review. Such ``fully-AI'' systems often exhibit high self-reported error rates and lack domain-specific skills, failing to address the particular challenges of experimental physics \cite{liu_autoresearch_2026,tang_fars_2026}. From a practical perspective, scientific infrastructure is built around human accountability and verification, and most journals and preprint servers only accept human authorship \cite{APS_AI_2026,COPE_AI_2023,arXiv_AI_2023,bioRxiv_AI}. Thus, real-world AI-assisted scientific work requires human researchers to deduce physical principles from AI outputs, test their validity and the range over which they hold, and connect them to independent experimental and computational evidence, resulting in a scientific finding that a human author can publicly endorse.

Inspired by these challenges and our extensive exploration of customized, closed-loop workflows, this paper serves as a practical tutorial on conducting collaborative, LLM-driven agentic research in experimental physics (fundamental and applied). Instead of focusing on a custom framework aimed at a single problem, we begin by formulating ten behavioral principles that we want a collaborative AI agent to follow in physics research, spanning scientific norms, effective agent use, and domain-specific knowledge (Sec.~\ref{sec:rules}). We then detail methods to implement these principles, including system prompts, skills, workflows, templates, and programmatic gates and structures for interactive and closed-loop work (Sec.~\ref{sec:framework}) with the goal of providing a guide that can be applied to a wide range of problems; these methods are instantiated in our research framework, PhysKit~\cite{reed_physkit_2026}, which we present as a concrete example. Finally, we present six case studies showcasing the capacity of structured agentic physics research across simulation, hardware-in-the-loop control, device design, symbolic reasoning, and neural network optimization (Sec.~\ref{sec:cases}). Among the results discussed below, a collaborative AI agent identified a photodynamic model of a spin defect in hexagonal boron nitride (hBN) that explains experimental data better than published models; discovered candidate quantum control sequences that improve simulated performance by more than an order of magnitude relative to published sequences; commissioned a nitrogen-vacancy (NV) diamond nuclear magnetic resonance (NMR) spectrometer from an uncalibrated state, reaching $\sim$1~Hz $^{1}$H NMR frequency precision; and reconstructed the per-island magnetization reversal of a 2,178-island artificial spin-ice (ASI) array from diffraction-limited magnetometry, an analysis now incorporated into a forthcoming experimental study. We conclude with a discussion of the impact of agentic work on our research, the role of human researchers (users) in AI pipelines, the importance of developing protocols for collaborative AI use in scientific contexts, and the future of AI in physics research (Sec.~\ref{sec:discussion}). 

\section{Principles for collaborative agentic research}
\label{sec:rules}

We aim to design an LLM-based agent that usefully functions  as a collaborating researcher. An experimental physicist brings to a research team both norms governing how they log, verify, and report results \cite{feynman_cargo_1974,mitroff_norms_1974}; and the judgment to make recurring research decisions with incomplete information, such as which physical model and approximations to adopt, what experimental protocols and techniques to employ, what data to gather and trust, whether a result survives dimensional, order-of-magnitude, limiting-case, and symmetry checks, when an apparatus requires calibration or debugging, and how a proposed approach to these and other challenges might fail~\cite{wieman_how_2022, price_detailed_2021}. While current LLMs encounter these norms and decision patterns throughout their training data and apply them capably on benchmark problems from physics olympiads \cite{yu_hipho_2025}, they do so far less reliably on open, research-level problems \cite{critpt_2025}, which demand greater judgment and resist benchmarking. As LLMs are designed to handle a wide range of tasks and contexts, the burden falls on the user to clearly communicate desired procedures, workflows, documentation norms, and outputs for a given project, something that even advances in AI reasoning skills cannot necessarily address. Therefore, we begin by carefully considering what behavior is desired from a collaborative AI research agent and what workflows, organizational structures, and tools best support scientific work, with a particular focus on experimental physics at the benchtop scale.

\subsection{Scientific Norms}
\label{sec:pillars-norms}

\pillarhead{1. Honesty}{Report findings honestly, including weaknesses, caveats, and negative results, and never manipulate a metric.}

LLMs are known to exploit objectives \cite{krakovna_specification_2020, denison_sycophancy_2024, lu_ai_2024, helff_llms_2026}, hallucinate numbers \cite{lu_ai_2024, beel_evaluating_2025, liu_autoresearch_2026} and citations \cite{naser_how_2026}, fabricate claims \cite{schwartz_vibe_2026}, and misjudge quantitative comparisons \cite{lu_ai_2024, li_exposing_2025}. Therefore, collaborative agentic research workflows should emphasize the value of accurate results (positive or negative), freeze key metrics so they cannot be exploited, and require that numerical claims be traceable to executed code rather than directly computed by agents.

\pillarhead{2. Verification}{Use structured procedures, unit tests, and independent subagents to validate results before accepting them as correct.}

To further reduce hallucinations and false claims, collaborative agentic research should include multiple layers of verification that can detect coding, physics, or logic errors, especially as tasks become more complex. Verification should occur after every key step and should consist of programmatic unit tests, when appropriate, as well as specific checks assessing uncertainty, physical plausibility, appropriate dimensions and units, and noise features. Since an agent's stated reasoning can be biased, and LLMs assess their own reasoning poorly \cite{turpin_language_2023,chen_reasoning_2025,huang_large_2023} in critical situations, verification should be handled by independent subagents blind to the reasoning the first agent used. Using different LLMs for initial work and verification further mitigates blind spots \cite{lu_when_2025}. In order to enable further verification by a human researcher, the collaborative AI agent should record input parameters and assumptions in a standardized format, save all code in an organized structure, and produce relevant visualizations and reports.

\pillarhead{3. Methodical Inquiry and Traceability}{Decompose a problem into individually checkable steps, and record each step, its inputs, and its source.}

To improve the consistency of agent behavior, creative and reasoning-based steps should be embedded within predefined structures that enforce \textbf{honesty} and \textbf{verification} (see above), enable scientific practices such as citation and code availability, and build basic infrastructure that domain expertise can iterate on (see Sec.~\ref{sec:pillars-domain}). A provenance system should record the source of figures and claims, whether they come from existing literature, user-provided information, direct measurements, model-based simulations, or derived quantities. All code should be well-documented and version-controlled, and provenance records should document the specific version of the code that produced the figure or claim. Derivations should be clearly noted and preserved. Workflows should enforce planning and verification steps; and step-by-step procedures should ensure that common domain-specific concerns are addressed, including calculating uncertainty, considering geometric and other physical constraints, or comparing the answer to trivial solutions. 

\subsection{Effective Collaborative Agent Use}
\label{sec:pillars-collab}

\pillarhead{4. Communication}{Specify the agent's desired behavior explicitly and falsifiably, and include the rationale, stakes, and context for every task and handoff.}

Language used to communicate with research agents must be chosen with care, as it measurably affects agent behavior \cite{schulhoff_prompt_2024}. General guidance and desired workflow should be contained within scaffolding documents and system prompts, allowing the user to focus on task-specific prompts. Scaffolding documents should state desired behavior explicitly, with an emphasis on concrete actions against which the agent can objectively evaluate its compliance. The agent should seek clarifying information from the user rather than making assumptions, and the system should aid in the creation of scaffolding documents for complex projects, reducing the burden on the user to create them from scratch. 

\pillarhead{5. Focus}{Keep the agent's working context small and focused on the task at hand.}

Context is the full set of input text (including system and user prompts, conversation history, and retrieved data) that an agent reads to generate a response. Unnecessary context adds to the token cost and degrades performance: models retrieve mid-context information less reliably \cite{liu_lost_2023}, irrelevant material lowers accuracy even when the needed information remains present \cite{shi_large_2023}, and working loops degrade as prompts grow and orienting instructions fade \cite{lv_exploiting_2025, dongre_when_2026}. The context window should therefore be carefully managed. Scaffolding should be separated into tiers, with a persistent core of essential information accompanied by a larger body of supplementary documents ingested only during specific tasks. Longer sessions should make use of subagents or context compaction to maintain the most important information in the working context and avoid dilution with every reasoning step from a long session.

\pillarhead{6. Autonomy \& Safety}{Ensure user intervention is efficient and targeted.}

Collaborative agentic frameworks must prioritize both safety and efficiency. They should include mechanisms (gates) for detecting potentially harmful or irreversible actions and flagging them for user intervention or sign-off. For experimental physics, these gates should cover use of computational resources and file management, significant changes to the framework itself (see \textbf{Adaptability}), safe control of physical experiments, and researcher-defined criteria for data recording and processing, with thresholds adjusted based on risk severity. At the same time, closed-loop agentic work is a powerful tool, and early or excessive stops are a common failure mode. Therefore, the framework should allow variations of parameters within a safe range, and contain mechanisms to prevent early stops. Such mechanisms include embedding loops within a programmatic loop that handles calling new agents, eliminating the need for a single agent to continue working for extended periods. To take optimal advantage of user input, project goals and boundaries should be mutually agreed on by the user and collaborative agent before closed-loop sessions begin, giving the user an opportunity to provide targeted feedback and corrections in a single session and enabling productive iteration in longer independent runs. 

\pillarhead{7. Adaptability}{Keep the framework as a living document that improves through iteration.}

Collaborative agentic frameworks should be flexible and adaptable in response to user interactions, outcomes, new domains and use cases, LLM advances, and other software updates. In order to ensure revisions are governed with the same evidential stringency as results, clear rules and procedures should be created for editing and maintaining the framework to minimize unnecessary growth \cite{vanstein_code_2025, orlanski_slopcodebench_2026}, drift from its design principles, or overly specific changes \cite{fu_textreg_2026}. Improvements should be easy for the user to initiate, and standardized review of sessions and user feedback should identify areas for improvement. 

\pillarhead{8. Resource Management}{Use experimental and computational resources deliberately and efficiently, and account for their constraints.}

As collaborative agents are developed for a wide range of users and settings, they do not always properly take into account user-specific resource limitations such as the intended wall time of the campaign, local simulation capacity, cluster compute allocations, or experimental duty cycles and resource constraints. Therefore, the framework should include well-defined and tested instructions to ensure experimental and computational resources are used efficiently, including performing shorter tests before launching long runs, monitoring to catch failures early, and assessing the cost versus return when considering longer run times. Agents should be directed to consider the required computational and experimental resources (including wall time) when comparing protocols; and general procedures, including closed-loop iteration, should be both adaptable and user-regulated in order to accommodate experiments of different time scales and other resource constraints (e.g., peak power usage or consumption of finite experimental materials).

\subsection{Domain Expertise}
\label{sec:pillars-domain}

\pillarhead{9. Physics-Informed Reasoning}{Ground every step in the relevant physics, including geometry, units, scales, symmetries, and uncertainty, and judge results on physical plausibility.}

LLMs are fundamentally generalists, and while they excel in a wide range of domains including earning gold medals in physics olympiads \cite{yu_hipho_2025}, they remain far weaker on research-level physics problems \cite{critpt_2025,prlbench_2026} and can still return physically incorrect solutions \cite{schwartz_vibe_2026}. These limitations can be mitigated by incorporating domain-specific information into the framework that guides the collaborative AI agent. For example, supplying an LLM with retrieved, physics-relevant knowledge can improve its performance on physics problems~\cite{zheng_benchmarking_2025}. The same holds in agent-driven closed-loop discovery, where domain priors can improve results on real inverse-design problems \cite{yin_optimizing_2025} and physics-constrained search yields compact turbulence models that generalize to flow conditions outside the training data \cite{zhang_autoturb_2025}. For applications in experimental physics, domain-informed reasoning skills can include checking trivial cases, dimensional analysis, proper scaling of results with physical parameters, simulation validation and verification procedures, geometric and physical compatibility tests, uncertainty propagation, and statistical tests.

\pillarhead{10. Experimental Reasoning}{Employ hardware-in-the-loop protocols for experimental control tasks.}

Experimental tasks, including instrument and measurement protocol design, direct hardware operation, and data analysis, rely on specialized expertise and detailed technical information; hence, agents performing these tasks benefit from dedicated procedures and guidance. Unlike computational or analytic tasks where systems may behave deterministically according to known rules, experiments can include unknown behavior that must be observed empirically (e.g., instrument performance, environmental conditions). An experimental research framework should support reasoning about calibration, environmental drift, measurement time concerns, noise management, experiment characterization, hardware-specific debugging and control protocols, and experimental design practices. Agents often do not have context about what ``reasonable'' suggestions are for experimental improvements, so the user should set clear boundaries on feasible changes to experimental setups.  

\section{PhysKit: A Structured Environment for Collaborative Agentic Research}
\label{sec:framework}

\subsection{From Principles to Structure}
\label{sec:structure}

To transform off-the-shelf LLMs into reliable and collaborative research assistants that reflect the above principles, we developed PhysKit, a structured research framework implemented through a collection of ``skills'' for experimental physics. Here, a \textit{skill} consists of a Markdown file delineating in natural language a procedure for an AI agent to follow for a particular class of tasks, optionally accompanied by scripts, templates, and reference material. Several widely used LLM harnesses, including Claude Code and Codex, implement skills as a standard feature \cite{anthropic2026claudeskills,openai2026buildskills}. A user can invoke skills directly or they can be automatically triggered by the agent if the user's prompt matches a skill's description. While the systematic study of skills remains at an early stage, recent preprints report better average task completion and instructional compliance with customized skills \cite{li2026skillsbench,shaposhnikov2026agenticskills}. 

PhysKit directly encodes the principles for collaborative agentic research, outlined in Sec.~\ref{sec:rules}, into its skills and the resulting procedures, focusing on tasks pertinent to experimental physics research. Besides natural-language instructions, PhysKit also embeds deterministic programmatic components within agent-driven workflows. These components handle mechanically specifiable operations (including computing predefined metrics, running executable correctness checks, and recording standardized experimental products) while agents remain responsible for formulating hypotheses, interpreting results, and exercising scientific judgment. While no framework can completely eliminate the risk of error, in our experience PhysKit can substantially reduce time spent redirecting agents as well as the effort required to audit and correct their output. In Sec.~\ref{sec:cases} below, we describe six case-study applications of the PhysKit framework, drawn from the Walsworth group's active experimental projects. A precise quantitative assessment of the framework's performance and systematic effects remains for future work.

The current version of PhysKit consists of 26 skills, which can be grouped roughly into four categories: 
\begin{enumerate}
    \item \textbf{Campaign orchestration skills} establish the research environment, organize its central workflow, and maintain its continuity (see Secs.~\ref{sec:autocampaign}--\ref{sec:interactive}); 
    \item \textbf{General skills} strengthen scientific assurance and communication, for example, by subjecting results to independent adversarial checks (performed by the \textit{verify} skill), building a battery of reusable tests for properties including analytic limits, numerical convergence, and conservation laws (\textit{correctness-suite}), and writing scientific reports (\textit{report});
    \item \textbf{Specialist skills} provide procedures for tasks such as analytical derivation (\emph{derive}), numerical debugging (\emph{debug}), dimensional analysis (\emph{dimensional-analysis}), and comparison of competing physical explanations (\emph{model-selection});
    \item \textbf{Development skills} handle the improvement and maintenance of the framework itself, including \textit{campaign-review}, which reviews a completed campaign and suggests improvements to the framework based on the results; and \textit{improve-kit}, which guides agents through making updates according to the framework's design principles.
\end{enumerate}

A complete list of PhysKit skills and their functions, along with detailed documentation, is publicly available via GitHub~\cite{reed_physkit_2026}. As the framework is applied to a broader range of tasks, we expect to add new skills and expand existing ones.

While PhysKit skills can be invoked directly by the user, they are designed to be interconnected and to compose larger research procedures. The \emph{verify} skill provides a central example. Rather than operating only as a standalone command, it is called by other skills, including \emph{derive}, \emph{discovery-loop}, \emph{experiment-design}, \emph{model-selection}, \emph{fork-campaign}, and \emph{report} before a result is relied on or reported. \emph{Verify} in turn draws on other skills: it invokes \emph{dimensional-analysis} as an initial units and bounding check, uses the physical inventory and expected ranges established by \emph{frame-problem}, and treats the state of the \emph{correctness-suite} as evidence about the reliability of the underlying numerical machinery.

PhysKit skills are also designed to create a standardized set of documents and organizational structure, to help manage complex projects across multiple sessions. This structure includes an index of all files in the project, a record of engineering discoveries about the project, and a list of key system and measurement parameters. The \textit{handoff} skill helps agents preserve work and transfer knowledge from one session to the next. Finally, since closed-loop research modes involve similar scaffolding documents for each campaign, PhysKit includes templates for key files that the agent adapts to the specific problem during setup. The templates help ensure information is recorded consistently, maintain compatibility with programmatic elements, and guard against omitted steps. Records are also kept in standardized formats, improving the transparency of project documentation to both human and AI readers. 

PhysKit supports two work modes: closed-loop campaigns (Sec.~\ref{sec:autocampaign}), in which the agent iterates semi-autonomously toward a fixed goal with minimal human intervention; and interactive sessions (Sec.~\ref{sec:interactive}), in which the researcher collaborates with the agent in real time in an open-ended conversation. Because both modes share the same organizational scaffolding, it is straightforward to switch between them. An extended interactive session can define goals and constraints for a subsequent closed-loop campaign, and such a campaign can be paused for an interactive session for the researcher to probe and verify its findings and decide whether any adjustments in strategy or goals must be made.

\subsection{Closed-Loop Research Campaigns}
\label{sec:autocampaign}

\begin{figure}[tbp]
  \centering
  \includegraphics{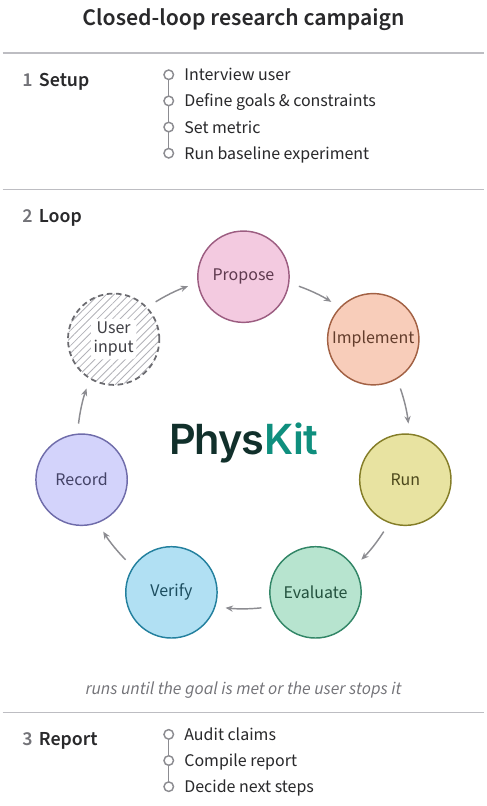}
  \caption{Life cycle of a collaborative agentic research campaign. After setting the goal with the human researcher (user), an AI agent repeats the loop on its own, then reports its results. The researcher sets the level of verification and can add ideas at any time.}
  \label{fig:loop}
\end{figure}

The closed-loop mode is designed for extended, semi-autonomous work toward a well-defined goal. During closed-loop campaigns, agents follow an iterative process, proposing, testing, verifying, and evolving solutions (Fig.~\ref{fig:loop}), aided by structured scaffolding documents that maintain project goals, rules, and results history. Unlike interactive sessions (Sec.~\ref{sec:interactive}), where the presence of the user allows immediate clarification and revision, a campaign must gather comprehensive details about the project's context and goals to allow it to run for long periods (hours or days) with minimal or no human intervention.

Work in closed-loop mode can be designed around the optimization of a single reward metric, as in finding a quantum pulse sequence with the best fidelity (see Sec.~\ref{sec:case-pulse}). However, it can also be centered on a well-defined, multifaceted goal, such as finding a physical model for experimental data that satisfies a list of criteria (e.g., low fit residuals, strong explanatory power for systematic effects, and consistency with published results). Finally, it is also possible to perform an open-ended ``discovery campaign'' that seeks to discover a new scientific product (a computer algorithm or mathematical expression, for example) fulfilling certain predefined criteria that are then rigorously validated (see Ref. \cite{romeraparedes_mathematical_2023} for a prior implementation in pure mathematics). 

\textit{Problem selection.} Not all problems are suitable for a full closed-loop research campaign. Such efforts provide the most value when the parameter space is large and poorly suited to conventional optimization techniques. In the pulse sequence design case study (Sec.~\ref{sec:case-pulse}), substantial physics-based reasoning is required to narrow down the enormous parameter space of all possible sequences into a few promising subtypes, which can then be addressed with exhaustive enumeration and/or conventional optimization algorithms. However, smaller-scale problems can often be efficiently solved through an interactive, user-attended session (Sec.~\ref{sec:interactive}). Since setting up the full campaign scaffolding requires a non-negligible upfront investment, a reasonable strategy is to evaluate a lightweight solution first through a single well-framed prompt before deciding whether to scaffold a full closed-loop campaign. PhysKit is designed to keep projects organized in a way that enables realistic switching between work modes as needed.

\textit{Defining the reward metric.} One of the most important elements of a productive long-term closed-loop campaign is defining precise reward metrics that correlate strongly with the target validity of the proposed solution. For some problems, this is fairly straightforward, because the metric itself is hard to satisfy by accident. For example, if the metric tests neural network performance on previously validated data (Sec.~\ref{sec:case-current}), a network that scores well is very likely coded correctly and trained on an appropriate dataset. In contrast, other problems have metrics that only correlate weakly with target validity. A common task is to fit a mathematical model to existing data (e.g., Sec.~\ref{sec:case-vb}), where the sum of squared fit residuals between the fit and data is the apparent metric. However, scientific judgment is required to decide whether the residuals reflect the statistical noise of the data or a systematic problem with the fit, the model, or the data. These problems often need secondary metrics or guardrails, such as tracking the parametric dependence of residuals for signs of systematic error or directly measuring fit accuracy on simulated data. Guardrail metrics are also suitable for problems that require multiple simultaneous constraints. One example is our microwave device design case study (Sec.~\ref{sec:case-mwdevice}), which has the goal of providing circularly polarized microwaves to a solid-state spin-defect sample, while meeting specified targets for experimental performance (peak microwave field strength and polarization purity) within instrument limitations, such as the maximum available microwave input power.

\textit{Campaign setup.} To start a new campaign with PhysKit, the user can directly invoke the \textit{new-campaign} skill with an appropriate prompt describing the task. If the initial prompt does not provide enough information, the skill launches a structured interview that asks the user to clarify the problem definition and framing; refine the campaign goal(s); define the constraints, physical model, and any software packages to be used; and set the target campaign duration. Here, it is important for the user to precisely and thoroughly describe the physical, experimental, and logistical constraints of the problem, point the agent to existing code and/or relevant literature, and carefully review the resulting campaign plan. Once the goals, metrics, and plan are approved by the user, they are recorded in the campaign's scaffolding documents. The goals and evaluator that computes the metric for each solution are then ``frozen'' so that the agent cannot game or modify them. Before iteration starts, the evaluator is tested on known controls and a baseline experiment validates the full pipeline using a known configuration and result. These procedures ensure that the evaluator is sound and provide a starting reference point against which subsequent solution candidates can be compared.

\textit{Running the campaign.} During each loop iteration of the closed-loop campaign, the agent is instructed to consult the campaign scaffolding documents and records; propose and implement modifications, evaluated using the frozen metric; interpret the outcome; and write new records about the result and what was learned. If an experiment appears to have succeeded in improving on the reward metric, it is recorded as the ``current best'' result together with a label stating how it was verified, and an agent ``hook'' (a script that fires automatically at a defined point in the agent's workflow~\cite{anthropic_hooks}) ensures that the label is accurate and that a refuted result is never promoted. The level of verification, which can include unit tests, self-review, or review by an independent agent, is determined by a user preset. If progress appears to stall, the agent follows PhysKit's guidance for changing approach. The loop runs continuously until the user intervenes, the agent determines that the campaign goal has been reached, or an optional predefined maximum campaign duration is reached.

Typically, closed-loop research campaigns on challenging tasks in experimental physics last anywhere from a few hours to days. In our case studies described below, the pulse sequence campaign ran 49 experiments in about 16 hours and scored roughly 10 million candidate sequences (Sec.~\ref{sec:case-pulse}); the NMR commissioning campaign ran 151 experiments in 20.8 hours against live hardware (Sec.~\ref{sec:case-nmr}); the microwave design campaign accumulated 418 full-wave candidate solutions over 140 solver-hours (Sec.~\ref{sec:case-mwdevice}); and the refit of optical data from boron vacancy quantum defects ($V_B^-$) in hBN tested 101 model structures over four days of largely unattended operation, with about 55 hours of fitting time (Sec.~\ref{sec:case-vb}). In such long campaigns, the user typically checks in periodically to track progress, redirect campaign strategy or subgoals when needed, and resolve occasional technical issues, interactions that often add substantial value compared to fully undirected iteration (see Secs.~\ref{sec:case-lessons} and \ref{sec:discussion} for more discussion of human intervention in closed-loop campaigns). 

\textit{Reporting.} A campaign ends with a final report that states the recommended solution and the decision it supports; documents the campaign history and scope, including runner-up candidates and their relative performance; tabulates every physical parameter with its provenance (literature, calculations, measurements) and the in-campaign evidence behind it; includes figures of key results; and carries a bibliography in which every entry has been checked against its real source. The report is itself verified by the agent before it reaches the researcher, with important values recomputed from the stored artifacts rather than transcribed. This last step is handled as a special case of the \textit{report} skill, which renders the document as compiled \LaTeX{} and records input parameters, assumptions, literature records, results, and conclusions in a step-by-step way for easy verification. Because every experiment within a campaign is archived with the exact code, configuration, pinned input data, and environment that produced it, any reported result can be traced, re-executed or reanalyzed, or reused as a starting point for future work.

\subsection{Interactive Sessions}
\label{sec:interactive}

In addition to closed-loop research campaigns for structured iteration, PhysKit supports a broader range of day-to-day research tasks through interactive sessions between the human user and collaborative agents, guided by general instructions and skills. Interactive sessions can be used for a variety of single-prompt tasks such as building a graphical user interface (GUI), processing a new dataset, or refining a fit. They can also refine the metrics, constraints, guardrails, and strategies required for a successful closed-loop research campaign. Interactive sessions require less setup than campaigns, as the user can provide context and corrections in real time, and are well suited to tasks that can be solved with minimal iteration, such as data visualization and modifications to existing workflows. Working within PhysKit sessions is also easier than working in an unstructured environment, as the framework handles general instructions on process, documentation, and verification, allowing the user to focus on task-specific prompts. The default workflow for interactive sessions is as follows:

\begin{enumerate}
  \item \textbf{Clarify.} Ensure the agent fully understands the user's goal and which decisions the work will inform, states the assumptions it is making, and raises the competing readings of an ambiguous request rather than choosing one without consulting the user. 
  \item \textbf{Frame.} Collaboratively plan how to approach the problem and how to know whether the answer is right, in the context of what is already known. This includes what is in and out of scope, which physical effects matter and which can be neglected, what the literature and the project's own records already establish, and what parameter range the answer should fall in.
  \item \textbf{Decompose.} Collaboratively break the task into steps that can be checked individually, and hand independent steps to subagents that work in parallel when appropriate.
  \item \textbf{Execute.} The agent carries out the steps above, keeping changes attributable, checking between steps where an error would compound, and propagating an uncertainty estimate through each quantity it derives.
  \item \textbf{Verify.} The agent validates results through unit tests and an independent subagent, working from a structured brief that targets specific error classes.
  \item \textbf{Record and Report.} The agent records and reports to the user what was done and what was found in the interactive session, including failures and the gap between predictions and the outcome, along with the configuration and code needed to reproduce the session.
\end{enumerate}

Each step is carried out by one or more skills that bring their own procedures and reference material, and can involve several substeps. For example, framing (step 2) calls the \textit{frame-problem} skill, which produces an inventory of the relevant physics with a stated reason for every effect left out, a verification plan written before any work is done, and a decomposition with checkpoints placed in proportion to risk. Verification (step 5) calls the \textit{verify} skill, which grades what it finds on a fixed scale rather than returning a binary verdict, so that a result resting on a weak premise is recorded differently from one that is simply wrong. The PhysKit core prompt also includes a list of general core principles that we find to be useful to mitigate some of the most common agent failure modes encountered through testing. These failure modes include a lack of honest and thorough disclosure of the limitations of results; prioritization of task ``success'' over accuracy; inadequate consideration of physical plausibility; weighing a claim's origin (including user errors) instead of evidence; and a tendency to treat technical creativity as incompatible with rigor.

\subsection{Hardware in the Loop}
\label{sec:hw}

PhysKit supports closed-loop research campaigns involving control of laboratory equipment, in which the agent designs, runs, and analyzes experiments with precautions taken to ensure safe operation. During campaign setup, the user and agent define the allowed range of every instrument setting, with hardware safety limits set at sustainable operating points rather than absolute maxima. Each proposed experiment then passes through a deterministic gate that rejects any proposal that omits a required setting (rather than filling in a default value) or that specifies a setting outside its allowed range, allowing the agent to correct the error before any instrument time is spent.

Approved experiments are handed to the laboratory's existing deterministic (non-agentic) instrument control program through an exchange of  files, so that either program can crash and restart mid-campaign without loss of information, and a full record of each request and its result remains on disk. The campaign loop monitors the experiment so it can intervene if an instrument stops producing data or fails to meet predetermined criteria. Each completed acquisition passes a data-quality check before it is scored by the campaign's frozen evaluator, after which the campaign proceeds as described in Sec.~\ref{sec:autocampaign}. The NV-NMR case study described in Sec.~\ref{sec:case-nmr} was conducted as such a campaign. 

\section{Case Studies}
\label{sec:cases}

Drawn from our laboratory's active projects, the six case studies below demonstrate the potential of agentic research to advance objectives in experimental physics. The closed-loop research campaigns outlined below (i) reduced the test error of a published current density reconstruction network by 33\%; (ii) produced a photodynamic model that predicts experimental data better than published models; (iii) designed a fabrication-ready microwave device for zero-field $V_B^-$ magnetometry; (iv) identified quantum control pulse sequences that outperform published sequences in simulation by more than an order of magnitude; (v) commissioned an NMR spectrometer to $\sim$1~Hz precision on live hardware; and (vi) built a fitting pipeline that recovers the magnetization state of each nanomagnet in an artificial spin-ice array. We deliberately select case studies within our own expertise in quantum sensing, but the methods themselves are not subfield-specific, and we expect them to generalize to other domains. 

\subsection{Current density reconstruction}
\label{sec:case-current}

\begin{figure}[!t]
  \centering
  \includegraphics[width=\columnwidth]{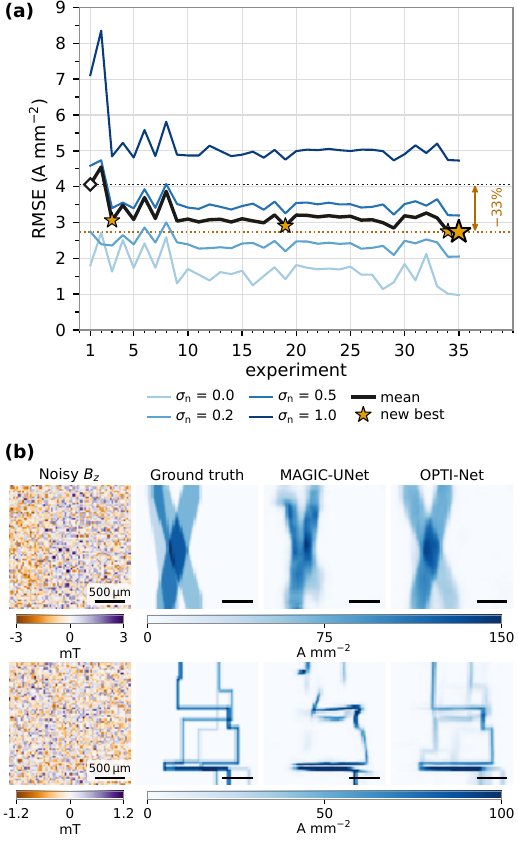}
  \caption{Current density reconstruction campaign. (a) Primary metric (root mean square error, RMSE) for the collaborative agentic campaign. RMSE is calculated from the difference between the ground truth and model prediction based on a fixed set of synthetic test data for different noise levels ($\sigma_N\sim$ inverse SNR), which also served as secondary guardrails. The previously developed MAGIC-UNet baseline \cite{reed_machine_2025} is indicated by a white diamond; and campaign iterations that achieved a new best are marked with gold stars. Candidates were confirmed with three seed runs and promoted only if they were significantly better than the variation due to seed spread ($<1\%$, too small to display in the figure). (b) Example current density reconstruction images from MAGIC-UNet and OPTI-Net (the optimal network resulting from this campaign) for the SNR = 1 case. From left to right, columns depict: (i) $B_z$ component of synthetic noisy vector magnetic field images (similar to those produced by a quantum diamond microscope~\cite{reed_machine_2025}) used as inputs; (ii) ground truth current density from the synthetic test data; (iii) MAGIC-UNet's current density reconstruction; and (iv) OPTI-Net's current density reconstruction on the same data.}
  \label{fig:case-current}
\end{figure}

Recovering the two-dimensional (2D) source current density spatial distribution from a measured 2D magnetic field map is a common inverse problem across physics, engineering, and biomedicine. The Biot–Savart law provides a simple forward calculation (from 2D current density distribution to 2D magnetic field pattern), but its inversion by traditional analytic methods amplifies noise without bound \cite{rothUsingMagnetometerImage1989}, severely degrading performance on low signal-to-noise ratio (SNR) data and motivating alternate approaches such as neural networks.

We used a closed-loop agentic workflow in PhysKit to search for improved neural network architectures relative to the previous best result, a hand-optimized convolutional neural network (CNN) developed for current density inversion (MAGIC-UNet) \cite{reed_machine_2025}. The agent was asked to reduce the average root mean square error (RMSE) between the ground truth and prediction on a fixed set of synthetic test data at four different noise levels. Before beginning the optimization, the agent ensured it could reproduce the paper results~\cite{reed_machine_2025}, which applied MAGIC-UNet to current density reconstruction from two-dimensional images of vector magnetic fields acquired by a quantum diamond microscope.  The agent corrected some subtle differences and audited saved models from the original work, until it was able to reproduce the paper's published structural similarity index measure (SSIM) table.

The agent explored a large network architecture and training configuration parameter space with a coarse-to-fine methodology, first investigating neural network architecture alternatives (Fourier neural operators \cite{li_fourier_2021}, shared-decoder and width-scaled U-Net variants \cite{ronneberger_unet_2015, zagoruyko_wide_2016}, symmetry-constrained networks \cite{cohen_group_2016, weiler_general_2019}); then pursuing training-procedure refinements of the winning family ($D_4$-equivariant U-Net); and finally testing hyperparameter changes (learning rate and schedule, optimizer, loss function, weight averaging, and training-noise distribution). The resulting optimal network (OPTI-Net) combines three complementary modifications to MAGIC-UNet~\cite{reed_machine_2025}: variable training-noise augmentation (the original network was trained at a single fixed noise level); exact equivariance under the dihedral group $D_4$ of the square pixel grid~\cite{dresselhaus_group_2008},\footnote{Although the reconstruction is posed on a 2D grid, $\mathbf{B}$ and $\mathbf{J}$ are 3D axial and polar vector fields, respectively, which carry different representations of $D_4$ than 2D image channels and must be transformed accordingly~\cite{jackson_classical_1999}.} enforced by passing all eight rotations and reflections of each input through a weight-shared U-Net, mapping each output back to the original frame, and averaging~\cite{yarotsky_universal_2022,basu_equituning_2023}; and cosine learning-rate annealing~\cite{loshchilov_sgdr_2017}. $D_4$ equivariance also allows OPTI-Net to use one-quarter as many trainable parameters as MAGIC-UNet (3.9 vs.\ 15.5 million), as the campaign found that a $D_4$-equivariant U-Net with half as many channels per layer outperforms full-width U-Nets without equivariance. Together, these changes reduce the OPTI-Net RMSE on synthetic test data by about 33\%  from the paper baseline (Fig. \ref{fig:case-current}a), with other metrics of image quality such as SSIM improving as well. The performance increase is greatest on noisy data (SNR $\sim$1), as seen in the reduction of spurious artifacts from noisy signals (Fig. \ref{fig:case-current}b), and quantified by the large increase in SSIM on SNR $\sim$1 data from about 0.72 for MAGIC-UNet to 0.90 for OPTI-Net. 

The agent designed the research campaign to use GPU time efficiently on a single 12~GB consumer GPU. A tiered protocol of short feasibility probes, epoch-capped single-seed screening runs, and full-length runs reserved for promising candidates helped ensure efficient use of GPU time; and the agent ultimately spent approximately 110~GPU-hours across 38 logged experiments. Because single-seed comparisons can promote favorable random initializations rather than better architectures (a phenomenon known as the ``winner's curse''), the agent measured the seed-to-seed spread of the benchmark directly, only promoting ideas if they clearly surpassed the seed spread and confirming winners with multiple seeds. 

The low marginal cost of synthetic experiments during the campaign allowed the user and agent to explore a series of physics-informed modifications to the OPTI-Net network and data-processing pipeline, including projecting the input field onto the divergence- and curl-free subspace admissible in a source-free measurement region (a provable denoising of the input); Wiener-style spectral preconditioning~\cite{wiener_extrapolation_1949} built from the known field-to-current transfer function; and a current-continuity penalty on the network output. Ultimately, none of these approaches improved current density reconstruction accuracy, suggesting that the OPTI-Net network was already successfully inferring the physics these other approaches are built on. In particular, when testing projections to a divergence- and curl-free  subspace, the agent found that MAGIC-UNet family models already attenuate nonphysical input components by a factor of $\sim$5--10, leading to almost no change in outputs after the projection despite a nominal increase in SNR.  This negative result provides a useful supplementary finding, which helps to explain the advantage of this class of network on noisy data over analytic methods, which do not exploit the additional constraints imposed by requiring only physically plausible magnetic field inputs \cite{rothUsingMagnetometerImage1989}.

\subsection{Improved photodynamic model for \texorpdfstring{$V_B^-$}{VB-} spin defects in hBN}
\label{sec:case-vb}

\begin{figure*}[tp]
  \centering
  \includegraphics[width=\textwidth]{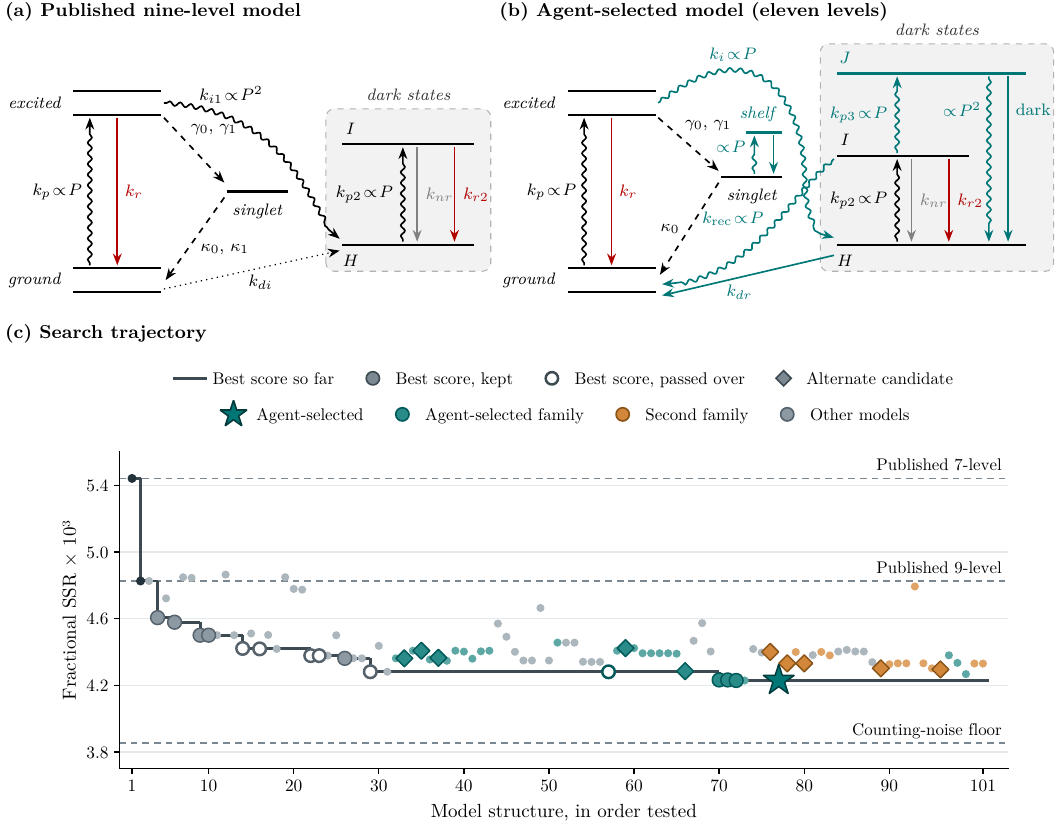}
  \caption{Campaign to optimize the photodynamic model for $V_B^-$ defects in hBN. In the energy level diagrams, each triplet state is drawn as its $m_s{=}0$ (lower) and $m_s{=}\pm1$ (upper) sublevels; wavy arrows are laser-driven transitions ($\propto P$ or $\propto P^2$), straight arrows are spontaneous decays, red (gray) arrows are radiative (nonradiative) decays, dashed arrows are intersystem crossings, and the dotted arrow is the dark conversion $k_{di}$. \textbf{(a)}~Manually formulated nine-level model, with ground and excited spin triplets, a metastable singlet, and a two-level dark manifold reached by photoconversion from the excited state ($k_{i1}\propto P^2$); published in Ref.~\cite{escalante_direct_2026}. \textbf{(b)}~Agent-selected model identified in the campaign. Model structures retained from the nine-level model ``baseline'' are black and additions or changes are teal. Key features include (i) photoconversion from the excited state as a one-photon step ($k_i\propto P$) rather than a two-photon step; (ii) a third dark state $J$; (iii) return paths from the dark states to the ground state, light-driven from $I$ ($k_{\mathrm{rec}}$) and spontaneous from $H$ ($k_{dr}$); and (iv) a light-driven shelf state attached to the singlet. The singlet's decay to $m_s{=}\pm1$ ($\kappa_1$) and the dark conversion $k_{di}$ are dropped. \textbf{(c)}~Best-fit fractional SSR for each of the 101 agent-investigated model structures, in the order tested (lower is better). The fractional SSR, the score used in the search, normalizes each unsmoothed PL trace by its own summed squared counts, so that all five laser powers contribute equally. The solid thick line tracks the best score so far. Marker shape shows the agent's decisions. Circles indicate models that held the best score up to that point in the campaign and were either kept as the preferred model (filled) or passed over due to poorer predictions of withheld laser powers or physical inconsistency (open). Diamonds are alternate candidates, kept despite a higher score because they better predicted withheld powers or were more physically consistent. The star is the final agent-selected model; and small points (of all colors) indicate scores that were not shortlisted. Color shows the set of states: teal, the same states as the agent-selected model (as in b); orange, the second model family discussed in the text; gray, all other models. Dashed lines mark the published seven- and nine-level models~\cite{escalante_direct_2026} (the first two points) and the PL counting-noise floor. Note that SSRs quoted in the text follow the convention of Ref.~\cite{escalante_direct_2026}: fits to smoothed PL traces with one shared normalization.}
  \label{fig:case-vb}
\end{figure*}

The negatively charged boron vacancy ($V_B^-$) spin defect in hBN is a leading candidate for nanoscale quantum sensing housed within a 2D van der Waals material~\cite{gottscholl_initialization_2020,vaidya_quantum_2023}. A spin-dependent intersystem crossing through a metastable singlet state enables $V_B^-$ optical spin polarization and spin-dependent photoluminescence (PL). Energy level transition rates are often extracted by fitting time-resolved PL measurements from $V_B^-$ ensembles to a seven-level rate-equation model or its reduced five-level form~\cite{baber_excited_2022,whitefield_magnetic_2023,cluaprovost_spindependent_2024,lee_intrinsic_2025}. However, the transition rates extracted with this methodology disagree substantially: e.g., reported singlet-state lifetimes at room temperature range from about 13~ns to 32~ns~\cite{lee_intrinsic_2025,baber_excited_2022,cluaprovost_spindependent_2024,whitefield_magnetic_2023}, whereas ab initio theory predicts the lifetime to be $\sim$seconds~\cite{reimers_photoluminescence_2020}. In recent work, our group directly measured the $V_B^-$ singlet-state lifetime to be 15(3)~ns, via time-resolved PL measurements enabled by a fast rise-time ($\sim$nanosecond) 515~nm laser~\cite{escalante_direct_2026}. We also formulated a nine-level photodynamic model, with two laser-power-dependent dark states (Fig.~\ref{fig:case-vb}a), that fits time-resolved PL data better than the previous seven-level model, including the observed PL quenching at high optical excitation power.

We used a closed-loop agentic workflow in PhysKit to search for improvements to the nine-level photodynamic model. The search fit to experimental data from Ref.~\cite{escalante_direct_2026} (five PL time traces for thermal and optically polarized $V_B^-$ spin ensembles at laser powers from 3.71 to 21.30~mW). The evaluator was frozen before the search; it fit each candidate model to all five PL traces as a set and scored the fit by the sum of squared residuals (SSR), with masking of the low-count bins between pulses and normalization such that all traces contribute equally, regardless of brightness (denoted fractional SSR in Fig.~\ref{fig:case-vb}c). Ref.~\cite{escalante_direct_2026} instead normalizes all five traces together and quotes the SSR against smoothed traces, which suppresses counting noise; we use that convention when comparing with the published fits. The agent could edit only the model specification: namely the number of energy level states, the transition rates between them, the laser-power scaling of each rate, which rates are fixed or free, and the spatial profile of the optical excitation across the flake. The radiative rate ($k_r$) and the $m_s{=}0$ intersystem-crossing rate ($\gamma_0$) were fixed to the values used in Ref.~\cite{escalante_direct_2026}, and the other fitted rates were checked against direct measurements in the literature for physical consistency. The campaign first refitted both published models (seven- and nine-level~\cite{escalante_direct_2026}) as baselines; refitted with all five traces normalized together, as in Ref.~\cite{escalante_direct_2026}, they reproduce the published fits. Because adding states increases the number of free parameters, a lower SSR alone does not establish a better model. To guard against overfitting and test the model's predictive power, promising models were refitted five times, once with each laser power withheld, and scored on their predictions of the withheld traces (leave-one-power-out cross-validation, LOPO).

The agent tested 101 model structures in 198 logged experiments over four days, accumulating about 55 hours of active fitting time. Every structure was fit from multiple sets of initial conditions and assessed by fractional SSR; promising structures were also scored on predictions for withheld laser powers. The best model emerged through four attributable additions to the published nine-level model: photoconversion out of the excited state as a one-photon step ($\propto P$) instead of a direct two-photon step ($\propto P^2$), together with light-driven recovery from the dark states; a third dark state; a light-driven shelf state attached to the singlet; and a return path from the dark states to the ground state in the dark, which lets the repeated pulse measurement reach a steady state. A final refinement, averaging the PL over the laser spot on a finite flake, improves the fit only slightly. 

The resulting agent-selected eleven-level model (Fig.~\ref{fig:case-vb}b) fits the full PL dataset better than both published models. Scored as in Ref.~\cite{escalante_direct_2026}, its SSR is 0.83, versus 2.13 for the nine-level and 3.44 for the seven-level model; scoring the unsmoothed traces instead adds about 4.6 to every model from counting noise and leaves the ranking unchanged. These values come from refits with all five PL traces normalized together, as was done for the published fits; the fractional SSR used in the search gives the same ranking (Fig.~\ref{fig:case-vb}c). The agent-selected model also predicts PL data at withheld laser powers much better than the seven- and nine-level models, with the clearest improvement over the nine-level model at the highest power (21.30~mW). Importantly, the agent did not simply follow the best fractional SSR score. It passed over six models that set a new best score, five because they predicted withheld laser powers worse and one for physical inconsistency, and it identified ten candidates with higher (worse) fractional SSR despite having better withheld laser power predictions or physical consistency (Fig.~\ref{fig:case-vb}c). After the campaign, the agent-selected model was fitted jointly to the canonical data and to two further datasets acquired in the study of Ref.~\cite{escalante_direct_2026}. It fit all three substantially better than the nine-level model (see appendix). 

The campaign also identified a second family of fourteen-level models that fit the PL data almost as well (SSR 1.00--1.10 in the convention of Ref.~\cite{escalante_direct_2026}, versus 0.83; orange diamonds at the right of Fig.~\ref{fig:case-vb}c). These models add the higher $^3E'$ triplet predicted by ab initio theory~\cite{reimers_photoluminescence_2020,ivady_ab_2020}, with part or all of the absorption and all of the photoconversion to the dark states routed through the $^3E'$ state, and implement a decay cascade through two singlet states instead of a shelf. Notably, both model families share certain changes to the nine-level model: one-photon entry into a three-state ladder of dark states, and paths from that ladder back to the ground state (see appendix). Models in this family show greater predictive power on LOPO tests under fractional SSR, which is why the agent kept them as alternates. Additionally, agents noted two cautions about the winning eleven-level model. First, its fitted intersystem-crossing ratio $\gamma_1/\gamma_0=2.66$ is slightly above the range of 2.0--2.5 implied by directly measured excited-state lifetimes~\cite{baber_excited_2022,cluaprovost_spindependent_2024,lee_intrinsic_2025}. Second, the light-driven shelf and the third dark state have no counterpart yet in ab initio theory. The alternate fourteen-level model family avoids both issues. 

Further experimental testing can distinguish between both candidate models and the original seven- and nine-level models through their differing predictions of PL saturation behavior at high laser power and recovery behavior with varied dark time. Agentic campaign mechanics, fitted parameters, and the detailed fitting procedure are given in the appendix.

\subsection{Microwave device design}
\label{sec:case-mwdevice}
\begin{figure*}[!t]
  \centering
  \includegraphics[width=\textwidth]{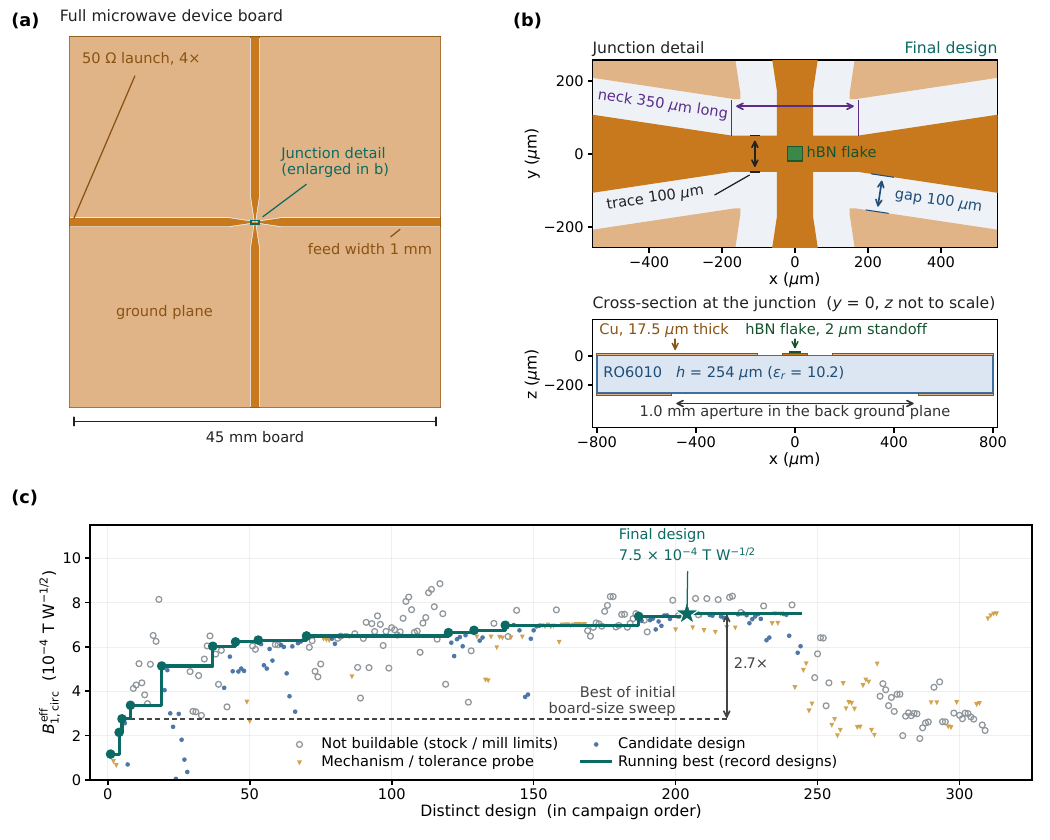}
  \caption{Microwave device campaign. \textbf{(a)}~Schematic of the final microwave device board design, produced by the collaborative agentic campaign: a 45~mm copper cross; the green box marks the junction region enlarged in (b). \textbf{(b)}~Top, copper at the junction: dark copper is the driven trace, pale copper is the ground plane, and light corridors are the milled clearance. Purple bars mark the 350-$\mu$m-long constant-width neck, and the green square is the hBN flake. Bottom, the layer stack, with a 1.0~mm aperture in the back ground plane beneath the flake. \textbf{(c)}~Calculated mean amplitude of the circularly polarized microwave field at the hBN flake at $\sim$3.5~GHz, per square root of amplifier power, for the microwave device designs generated during the campaign; the 418 simulations correspond to 313 distinct designs. The green line tracks the running best design, with a dot at each record and a star at the final design. The dashed line marks the best design from the agent's initial sweep of board size, a factor of 2.7 below the final design. Open circles are designs that could not be built (a substrate that is neither orderable nor on hand, a feature below the milling limit, or an idealized matching network); gold triangles are deliberate off-optimum probes of the aperture mechanism, fabrication tolerances, and mesh convergence. Neither is eligible to be the final design.}
  \label{fig:case-mwdevice}
\end{figure*}

Near-zero-field magnetometry with optically probed solid-state spin defects leaves the sample unperturbed and enables applications such as zero- to ultralow-field (ZULF) NMR \cite{blanchard_zulf_2016, zheng_zerofield_2019,lenz_zerofield_2021}. But at near-zero field, the two electronic spin resonances of an $S=1$ defect, $m_s=0\!\to\!\pm1$, overlap within the transition linewidth; hence linearly polarized microwaves, which drive both resonances, produce no first-order change in the optically detected magnetic resonance (ODMR) spectrum \cite{zheng_zerofield_2019} (see appendix).
 In contrast, circularly polarized microwaves preferentially drive one spin resonance and restore the first-order ODMR response in near-zero field \cite{alegre_polarization_2007, mrozek_circularly_2015,zheng_zerofield_2019,lenz_zerofield_2021}. However, experimental generation and delivery of circularly polarized microwave fields typically require two orthogonal components driven $90^\circ$ out of phase; and for $V_B^-$ defects, whose spin axis is normal to the hBN flake, both microwave components must lie \emph{in} the plane of the flake. Sadi \emph{et al.}\ \cite{sadi_spinstate_2025} reported spin-state-selective driving of $V_B^-$ in a few-mT bias field using circularly polarized microwaves delivered with a cross-shaped microwave waveguide; but the published description does not contain a detailed design, so the device could not simply be reproduced in our laboratory. We therefore set out to explore whether a collaborative agent equipped with an electromagnetic solver can generate a customized, possibly improved waveguide design for delivery of circularly polarized microwaves to resonantly drive $V_B^-$ spin defects in an hBN sample at $\sim$3.5~GHz.

Using PhysKit, we froze an evaluator that scores the mean amplitude of the in-plane circularly polarized microwave field at the hBN flake per square root of available amplifier power, computed with a finite-element solver (Ansys HFSS) and taken at the worse (lower amplitude) of the two circular polarizations. Four constraints set by our laboratory hardware were enforced as automatic pass/fail checks: conductor and gap widths of at least 100~$\mu$m, limited by the in-house circuit-board mill; a footprint within 50.8~mm, constrained by the sample stage; no metal in the optical window above the hBN flake; and drive from a single 16~W amplifier, which sets the maximum input microwave power. Two further guardrails were a polarization impurity threshold and impedance match to the 50~$\Omega$ feed. The topology, substrate, and all copper dimensions of the waveguide were left free.

The closed-loop agentic campaign ran for six days, scoring 418 simulations in about 140 solver-hours, which correspond to 313 distinct designs. In periodic check-ins, we twice corrected the agent's understanding of physics against the literature; otherwise, the agent operated largely independently, including unattended stretches of up to 27 hours. The final waveguide design is a symmetric copper cross (Fig.~\ref{fig:case-mwdevice}a, b): four identical 100~$\mu$m traces in one copper layer meet at the hBN flake site, making the two orthogonal drive axes equally strong by symmetry. In simulation, the agent-optimized waveguide delivers circularly polarized microwaves at $\sim$3.5~GHz with a mean amplitude of about $7.5\times10^{-4}$~T/$\sqrt{\mathrm{W}}$, yielding a $V_B^-$ electronic spin Rabi frequency of $\approx$119~MHz at 16~W input microwave power (see appendix), with a wrong-polarization field fraction of $\approx$2\%. This Rabi frequency is 2.7 times higher than that of the best design from the agent's initial sweep of board size (Fig.~\ref{fig:case-mwdevice}c). It corresponds to a 4.2~ns $\pi$-pulse duration, shorter than those achieved in many prior $V_B^-$ experiments \cite{wu_power_2025, liu_coherent_2022,lee_intrinsic_2025,rizzato_extending_2023,gong_coherent_2023}.

The most heavily tested finding was an unexpected design gain: a $\sim$1~mm aperture in the back ground plane beneath the hBN flake raises the mean amplitude of the circularly polarized microwave field by between $\sim$3\% and 15\% relative to an otherwise identical solid-plane board. This gain holds, with no counterexample, for all 13 controlled board pairs at the 100~$\mu$m conductor width of the final design, which span two laminates and three substrate thicknesses; the effect is consistent with removal of an opposing image current in the ground plane. The collaborative agentic campaign also documented an important negative result. Four 50~$\Omega$ waveguide arms meeting at a point are intrinsically mismatched, reflecting 81\% of the forward microwave power under the specified operating conditions; and none of 4{,}896 matching networks reached the $-10$~dB specification. As a result, the final agent-specified fabrication package for the microwave device includes an isolator (the available-power metric already accounts for this loss). Once the agent had selected a final design and written its fabrication files, we directed it to resolve its open questions, and the agent performed deliberately off-optimum diagnostics that localized the aperture gain to the conductor width and mapped a trade-off between microwave field strength and polarization purity. We note that all results in this campaign are still based on simulations; manufacturing and experimental validation of the device designs are ongoing.

\subsection{Pulse sequence design}
\label{sec:case-pulse}

\begin{figure*}[htbp!]
  \includegraphics[width=\textwidth]{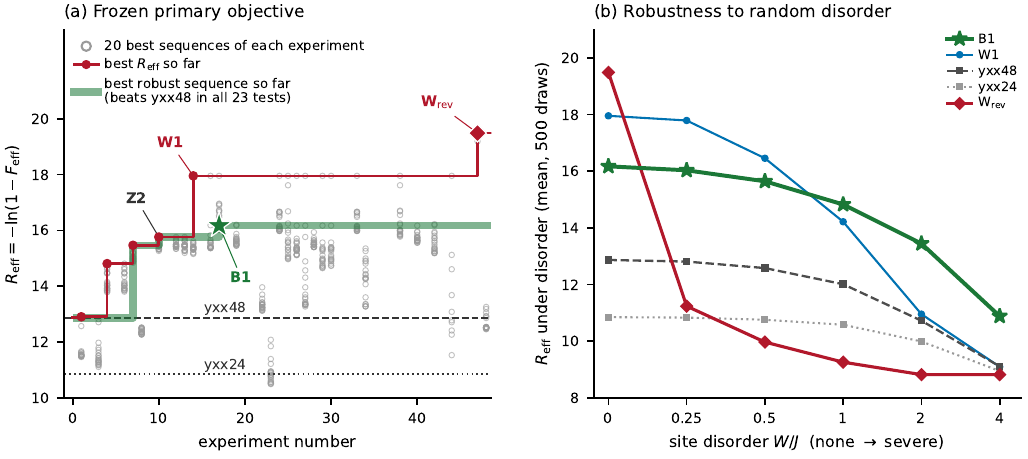}
    \caption{Pulse sequence design campaign, scored throughout by a frozen evaluator $R_{\rm eff}$. A total of 49 synthetic experiments were run in the 16-hour campaign. (a) The 20 best-scoring distinct sequences from each experiment (gray), with plots tracking the best $R_{\rm eff}$ thus far (red), and the best $R_{\rm eff}$ among robust sequences, i.e., those that beat the published benchmark (yxx48) at all 23 points of the robustness battery (green). Dashed and dotted lines mark the published benchmarks, and the horizontal axis gives the experiment number (one closed-loop iteration testing a single change). The final leading sequence W$_{\rm rev}$ (red diamond) beats the benchmark by almost three orders of magnitude in effective infidelity, whereas the best robust sequence B1 (green star) beats the benchmark by a factor of $\sim$30. (b) Robustness to random site disorder (a condition absent from the frozen evaluator): each spin receives an independent static detuning drawn uniformly from $[-W,+W]$, with $W$ in units of the dipolar coupling $J$, and $R_{\rm eff}$ is averaged over 500 random draws. W$_{\rm rev}$ falls below yxx48 already at $W=0.25J$, while B1 beats yxx48 at every disorder strength. Thus, while B1 never leads the best-$R_{\rm eff}$ staircase, it outperforms W$_{\rm rev}$ and other pulse sequences in a battery of robustness tests that probe likely performance in realistic experimental conditions beyond the reward metric (frozen evaluator) $R_{\rm eff}$.}
  \label{fig:case-pulse}
\end{figure*}

The coherence time ($T_2$) of solid-state spin systems, such as those commonly used in quantum sensing, is limited by environmental disorder and interactions with nearby spins~\cite{degen_quantum_2017}. Dynamical decoupling extends $T_2$ by interleaving free evolution with pulses that refocus the resulting spin phase accumulation~\cite{suter_protecting_2016}. In the simplest example, a Hahn echo sequence uses a single resonant $\pi$ pulse to reverse static phase accumulation. However, for a spin ensemble in which resonant spin–spin dipolar interactions become appreciable over the interrogation time, refocusing each spin is not enough. Global $\pi/2$ pulses must also be applied to rotate the secular dipolar interaction among different toggling frames, allowing both local disorder and spin–spin interactions to average toward zero over each sequence cycle. No general analytic method builds pulse sequences that satisfy this condition from first principles, even as average Hamiltonian theory (AHT)~\cite{brinkman2016aht} supplies some analytic conditions to narrow the search space. Thus, finding better pulse sequences to extend solid-state spin coherence times and improve quantum sensing performance remains an active area of research. Peng \emph{et al.} \cite{peng_deep_2022} addressed this challenge with deep reinforcement learning, producing an experimentally validated yxx24/yxx48 pulse sequence family that improved upon the well-known Cory48 sequence \cite{cory1990time}. We explored whether a closed-loop agentic workflow could improve upon the Peng \emph{et al.} result.

The collaborative agentic campaign adopted the same simulation setup as Ref.~\cite{peng_deep_2022}: a one-dimensional chain of three spins evolving under dipolar coupling $J$ between neighbors and a uniform frequency offset, with time discretized into steps that either idle or apply a single global $\pi/2$ rotation about the $\pm x$ or $\pm y$ axes. Systematic pulse angle error is included, so imperfections accumulate coherently and can only be removed by the ordering of the pulses, and not averaged away; and the frozen evaluator contains no noise or ensemble sampling and is exactly deterministic. We extended the maximum sequence length to 96 steps and normalized the score per step, so pulse sequences of different lengths are compared over equal total evolution time. The figure of merit (frozen evaluator) is $R_{\rm eff} = -\ln(1-F_{\rm eff})$, where the fidelity $F_{\rm eff}$ measures how close the state evolution is to the identity operation, so one unit of $R_{\rm eff}$ corresponds to a factor of $e$ (2.72) decrease in infidelity $1-F_{\rm eff}$. 

The campaign was set up with our PhysKit framework; and the agent was instructed to perform an open search for pulse sequences improving $R_{\rm eff}$ over the yxx48 sequence of Ref.~\cite{peng_deep_2022} without modifying the evaluator. Because a single metric is easy to overfit, every promoted candidate was additionally evaluated on a 23-point battery of robustness tests: sweeps of coupling-delay product, offset, and pulse angle, and also finite-pulse-width errors, applied in different combinations; and then the entire procedure was repeated on larger 8-spin, 10-spin, and 12-spin chains. The agent was connected to a local knowledge base of 89 indexed journal articles on relevant topics to inform the pulse sequence search. Unlike Ref.~\cite{peng_deep_2022}, we restricted the search to standard methods rather than machine learning methods, which tend to take longer per iteration and require a powerful GPU. The standard methods included exhaustive enumeration over structured subspaces, stochastic local search, and analytic constructions drawn from the literature.

After 14 synthetic experiments over roughly three hours, the campaign plateaued at Z2, a sequence scoring $R_{\rm eff} = 15.77$. A single, simple human instruction to consult the literature successfully broke the plateau, and the campaign reached $R_{\rm eff}=17.96$ with the sequence W1. This was the largest single gain of the campaign (Fig.~\ref{fig:case-pulse}a). Three experiments later, a variation of the same construction produced B1 ($R_{\rm eff} = 16.17$), which performs below W1 on the primary metric but beats yxx48 at all 23 points of the robustness battery. Further prompting across 25 more experiments did not improve on that success, and the agent reported that the remaining barrier was analytic rather than computational. We therefore had the agent compile its findings into a self-contained problem statement, which was passed to a separate reasoning model capable of sustaining long chains of mathematical reasoning. After an hour, the latter returned W$_{\rm rev}$ ($R_{\rm eff} = 19.49$), which was fed back into the main agentic loop and assessed with the same evaluator. The agent warned, however, that W$_{\rm rev}$ beat B1 in only two conditions across the 23-point robustness battery, suggesting that past a certain point, the main $R_{\rm eff}$ metric alone no longer strongly correlated with practically useful sequences. The campaign completed with 49 experiments over about 16 hours, scoring roughly 10 million candidate sequences. 

Following Ref.~\cite{peng_deep_2022}, we then tested the best sequences in a condition absent from the frozen evaluator: each spin receives an independent random static detuning drawn uniformly from $[-W,+W]$, imitating the inhomogeneous local fields of a real spin ensemble, and we compute the average $R_{\rm eff}$ from 500 runs. While the W$_{\rm rev}$ sequence has the highest value of $R_{\rm eff}$ for the main evaluation metric (reaching an effective infidelity of $3.4\times10^{-9}$, $753\times$ below yxx48, and keeping its lead up to a 12-spin chain), it performed poorly on the random disorder test, dropping below the performance of the yxx48 benchmark at modest disorder (Fig.~\ref{fig:case-pulse}b). B1 behaves in the opposite way: it beats yxx48 under all tested conditions, with 15--27$\times$ lower effective infidelity up to $W=2J$. We note that these results establish pulse sequence performance only under the simulated physical model; a full verification of performance requires an experimental demonstration, which is currently being planned.

\subsection{Nuclear magnetic resonance (NMR) experiment commissioning}
\label{sec:case-nmr}
\begin{figure}
  \centering
  \includegraphics[width=\columnwidth]{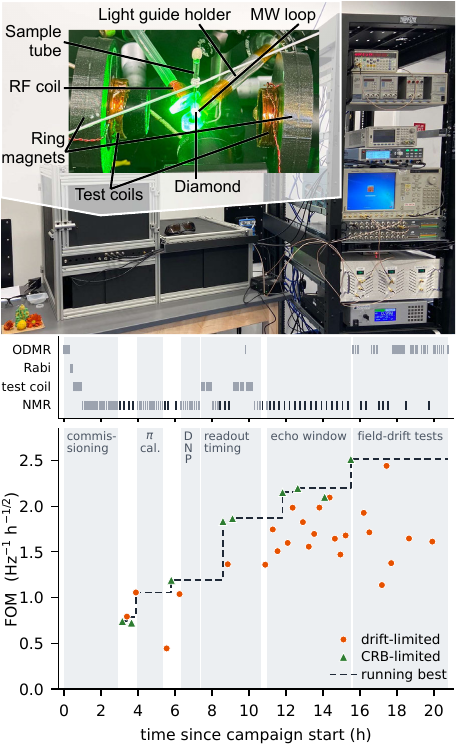}
  \caption{Apparatus and figure-of-merit (FOM) trajectory of the NV-NMR commissioning campaign (151 experiments over 20.8\,h). Top: photo of NV-NMR apparatus. Inset shows details of the experimental arrangement of the glass sample tube in contact with the diamond, which is in turn mounted on a liquid light guide to collect NV fluorescence. 
  A pair of ring magnets produces a 20.95 mT bias field, a shorted loop around the diamond provides microwave control of the NVs, and a tuned coil is used to apply rf pulses to 1H nuclei.
  Middle: every experiment against its start time, by type (NV ODMR, NV Rabi, test-coil injection of an RF signal, and NV-NMR); dark ticks mark the 34 scored NV-NMR precision experiments. Bottom: the FOM of each scored experiment at its completion time, marked by the term that set its uncertainty $\sigma_{\mathrm{arm}}$ [Eq.~(\ref{eq:nmr-sigma-arm})]: green triangles, the estimator's Cram\'er--Rao-bound uncertainty (``CRB-limited,'' Birge ratio $\leq 1$); orange circles, the observed block-to-block scatter of the $^{1}$H NMR frequency estimates (``drift-limited,'' Birge ratio $>1$). The dashed line is the running best. Shaded bands mark the agent's successive studies: commissioning (protocol steps 1--7, from NV ODMR through NMR pulse calibration), recalibration of the NMR $\pi$ and $\pi/2$ pulses, DNP pump settings, NV readout timing, the echo spacing and echo-train length, and repeated ODMR tests of whether the DNP pump shifts the bias field. The unshaded gaps between bands contain the scored experiments that linked successive studies.}
  \label{fig:case-nmr}
\end{figure}
NMR is a central analytical tool in chemistry, biology, and materials science. NV-NMR replaces a conventional pickup coil with NV color centers in diamond to enable NMR spectroscopy and imaging of micron-scale and nanoscale samples~\cite{devience_nanoscale_2015,lovchinsky_nuclear_2016,glenn_highresolution_2018,bucher_hyperpolarization_2020,arunkumar_micronscale_2021}. 
Setting up an NV-NMR measurement (locating the resonances, calibrating the pulses, and then optimizing signal acquisition) is currently a detailed, time-consuming task that requires an expert's focused attention~\cite{bucher_quantum_2019}. 
Thus, we tasked an agent, working within PhysKit, with commissioning an NV-NMR spectrometer from an uncalibrated state, and then optimizing its $^{1}$H NMR spectroscopy frequency precision within a 36-hour budget. 
The NV-NMR spectrometer operated with a bias magnetic field near 21\,mT and a sample of water (volume $<$1~nanoliter) containing 5\,mM 4-hydroxy-TEMPO for signal enhancement via Overhauser dynamic nuclear polarization (DNP)~\cite{bucher_hyperpolarization_2020}.
The agent was given control of the experiment hardware (Sec.~\ref{sec:hw}) and told only essential technical information about the apparatus: the NMR sample's identity; that the ensemble-NV-diamond sensor is $^{15}$N-enriched; and operational and safety limits on applied microwave and radio-frequency (RF) waveform amplitude, pulse duration, sweep length, and DNP pulse energy.
The agent was also given an eight-step commissioning protocol that fixes the order of the following tasks while containing no experimental parameters: (1)~broadband ODMR of the NV-diamond sensor; (2)~hyperfine-resolved NV ODMR; (3)~Rabi calibration for driving the NV ensemble electronic spins; (4)~NV-NMR detection-chain validation and optimization, using a test-signal coil to apply an RF signal similar to the expected NMR signal from the sample; (5)~setting the DNP frequency on a hyperfine transition of the radical; (6--7)~calibration of the NMR $\pi$ and $\pi/2$ pulses; and (8)~NV-NMR spectroscopy on the water sample, optimized under the frozen figure of merit.
The agent had to determine the bias magnetic field value, the NV and $^{1}$H spin resonance frequencies, the DNP frequency for the radical, pulse lengths and amplitudes, and the pulse sequence timing (echo spacing, number of echoes, acquisition window, and DNP duration) used for system characterization, NV-NMR spectroscopy, and DNP.
The campaign's frozen figure of merit (FOM) rewards time-normalized $^{1}$H NMR frequency precision. Each scored experiment acquires $n_{\mathrm{blocks}}$ consecutive blocks of NV-NMR data, and the estimator extracts an $^{1}$H NMR frequency from each block. The FOM is $1/(\sigma_{\mathrm{arm}}\sqrt{T})$, where $T$ is the wall-clock duration of the experiment in hours and
\begin{equation}
  \sigma_{\mathrm{arm}} = \frac{\max\left(\bar{\sigma}_{\mathrm{est}},\, s\right)}{\sqrt{n_{\mathrm{blocks}}}} .
  \label{eq:nmr-sigma-arm}
\end{equation}
Here $s$ is the observed block-to-block scatter (standard deviation) of the $^{1}$H NMR frequency estimates, and $\bar{\sigma}_{\mathrm{est}}$ is the mean per-block frequency uncertainty reported by the estimator, calibrated to the Cram\'er--Rao bound (CRB) for the measured SNR; the estimator excludes the first 0.5\,ms of each echo, during which the applied signal from the RF coil rings down. This FOM does not change when an NV-NMR spectrum acquisition is sliced into more or fewer blocks, does not reward (i.e., down-scores) scatter that happens to fall below the estimator's uncertainty, and penalizes measured NV-NMR spectral drift until the error model accounts for it. The ratio $s/\bar{\sigma}_{\mathrm{est}}$ is the Birge ratio\,\cite{birge_calculation_1932}: a value near 1 means the estimator's error model explains the observed scatter, and a value well above 1 signals unmodeled effects such as drift. We call experiments with a Birge ratio $\leq 1$ CRB-limited and those with a Birge ratio $>1$ drift-limited (Fig.\,\ref{fig:case-nmr}).

The campaign ran unattended for 20.8 hours, comprising 151 experiments (Fig.\,\ref{fig:case-nmr}). The agent completed protocol steps 1--7 in about three hours. From broadband and hyperfine-resolved ODMR it determined a bias field of 20.95\,mT and chose an ODMR spectral feature for NMR sensing, on which it calibrated a 116~ns NV $\pi$ pulse. With the test-signal coil, the agent showed that the detection chain accurately recovers an injected RF tone ($\approx$0.9~MHz) to within 0.04~Hz, at two different sensing frequencies and with the tone offset from the sensing frequency; and it measured an initial RF sensitivity of 258\,pT/$\sqrt{\mathrm{Hz}}$, already below the protocol's 300\,pT/$\sqrt{\mathrm{Hz}}$ target, which it improved to 162--177\,pT/$\sqrt{\mathrm{Hz}}$ by doubling the number of XY16 repetitions in the sensing sequence. The first NV-NMR $^{1}$H signal was detected at $t\,{+}\,61$~min, at 891.8\,kHz, and a pump-off control confirmed by $t\,{+}\,73$\,min that NV-NMR signal detection requires DNP of the sample; the agent's 638.1\,MHz DNP frequency agrees with the known TEMPO hyperfine structure. After calibrating the NMR pulses (RF pulses applied to the sample to induce a detectable NMR signal), it ran its first scored experiment at $t\,{+}\,3.2$\,h (FOM 0.74).

The agent then optimized one parameter at a time, bracketing each change with interleaved controls and replicates. It rechecked its NMR pulse calibration, the DNP pump settings, and the NV PL readout timing. Tuning the readout timing (the delay of the photodetector PL acquisition relative to the readout laser pulse) improved the test-signal sensitivity further, to 104\,pT/$\sqrt{\mathrm{Hz}}$, and raised the FOM from 1.19 to 1.87 at $t\,{\approx}\,9$\,h. The best acquisition delay, 0.8\,$\mu$s, agrees with the 0.6--0.9\,$\mu$s optimum we had previously found by hand on this apparatus. Finally, at a fixed echo-train duration of about 0.4\,s, the agent lengthened the echo spacing from 6 to 15\,ms (and shortened the train from 64 to 26 echoes). Each step lengthened the signal record per echo, which sharpened the estimator's frequency resolution and lowered $\bar{\sigma}_{\mathrm{est}}$, so the FOM that a CRB-limited experiment would reach rose from 1.86 to 2.60. The best experiment, at $t\,{+}\,15.5$\,h, reached FOM 2.51: a $^{1}$H NMR frequency uncertainty of 0.97~Hz in one 10-minute run on the 891.8\,kHz NMR spectral line, or 1.1\,ppm fractional precision. Across all ten experiments at these final settings, $\sigma_{\mathrm{arm}}$ ranged from 0.97 to 2.1\,Hz (median 1.47\,Hz).

What limited the ultimate NV-NMR frequency precision was system drift, not intrinsic NMR signal or NV sensor properties. Of the 34 scored experiments, 25 were drift-limited, as were nine of the ten at the final settings, with a mean block scatter of 4.5~Hz against a mean estimator uncertainty of 3.0\,Hz. The agent traced the excess scatter to the history of the DNP pump. After 30 minutes or more without pumping, the $^{1}$H line was 32--36~Hz higher, and the NV ODMR lines shifted by up to $\sim$50\,kHz, reflecting both a temperature change and a change of up to 0.7\,$\mu$T in the bias field at the diamond. The agent spent most of the final five hours on ODMR sweeps testing whether this pump-driven field change explains the scatter; the test was inconclusive. Having found no remaining parameter to adjust, the agent declared its search complete at $t\,{+}\,17$\,h, and we ended the campaign at $t\,{+}\,20.8$~h, with 15 hours of its budget unused.

An independent claim audit (Sec.~\ref{sec:autocampaign}) then checked the campaign's record. It rescored all 34 experiments from the raw data, reproduced the headline result in a fresh process, and confirmed the 638.1\,MHz DNP line against the radical's known magnetic parameters. It rejected the agent's claim of a replicated record: the best experiment was one of ten at identical settings and the only CRB-limited experiment among them, so we quote the median result as well as the best. The audit also found that the agent stopped lengthening the echo spacing while the FOM was still improving at the end of the tested range, identifying a direction for potential future performance improvement.

The experimentally realized NV-NMR frequency precision ($\approx$1~Hz per run requiring $\sim$10~minutes) is comparable to the $\sim$1~Hz spectral resolution achieved in high-resolution NV-NMR spectroscopy~\cite{glenn_highresolution_2018,smits_twodimensional_2019}. However, over the $\sim$20 hours of the campaign, the measured $^{1}$H NMR frequency moved by tens of hertz due to system drift, highlighting the opportunity for a future campaign to aid in improving system stability. The result shows that agentic research can be used successfully to commission a hardware experiment that pursues a sensitive precision measurement. For comparison, equivalent manual commissioning of the same NV-NMR apparatus takes multiple days of automated linear parameter sweeps or $\sim$10 hours of an expert's focused attention. The agent reached a scored NV-NMR measurement 3.2 hours after starting from an uncalibrated state, without supervision; and the same campaign can straightforwardly recommission the system from scratch after a bias field or NMR sample change.

\subsection{Artificial spin-ice (ASI) magnetization reconstruction}
\label{sec:case-asi}

\begin{figure*}
  \centering
  \includegraphics[width=\textwidth]{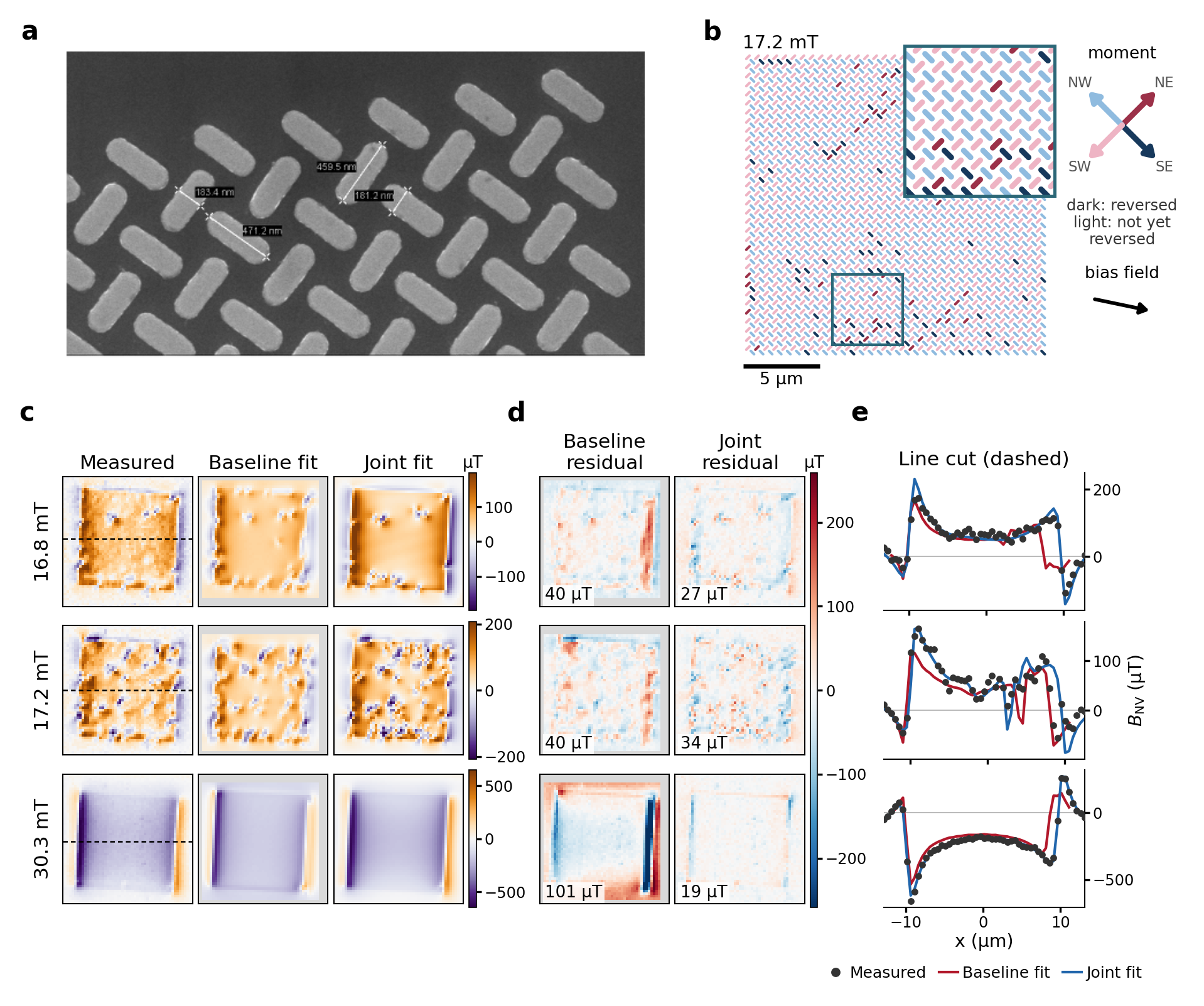}
  \caption{ASI magnetization reconstruction campaign. (a) Scanning electron microscope (SEM) image of a small example region of the pinwheel ASI; each nanomagnet island is $\sim$470~nm $\times$ 180~nm, with 2,178 islands in the entire ASI array. (b) Reconstructed microstate of the ASI under the final ``joint fit'' procedure at a bias field of 17.2 mT, early in the magnetization reversal (6\% of the array has reversed at this field). Each color corresponds to one in-plane moment direction (compass key). Dark colors (NE, SE) mark islands whose moment has reversed to point along the bias field, and light colors (NW, SW) mark islands that have not yet reversed. The black arrow shows the in-plane component of the bias field and the magnetic field projection of the NV confocal magnetometry data. The inset shows a magnification of the boxed region. (c) Measured NV magnetic field, $B_{\mathrm{NV}}$, after bias field subtraction, compared with the fields predicted by the baseline fit and the joint fit, at bias fields of 16.8 and 17.2 mT (early reversal) and 30.3 mT (saturation, all islands reversed). Gray regions are outside the field of view of the baseline fit. (d) Residuals (measured minus fit) for the baseline and joint fits. (e) Line cuts along the dashed line shown in (c) comparing the measured data and both fits.}
  \label{fig:case-asi}
\end{figure*}

Artificial spin ice (ASI) is a lithographically patterned, typically two-dimensional array of interacting single-domain nanomagnet islands (Fig. \ref{fig:case-asi}a), each behaving as a two-state Ising spin and arranged so that geometric frustration prevents their interactions from all being minimized at once \cite{skjaervo_advances_2020, li_superferromagnetism_2018}. ASI arrays are designer platforms for studying frustrated magnetism, emergent magnetic monopoles, and neuromorphic computing, all of which require reading out the microstate (the orientation of every island) as the array is driven through a magnetization reversal via an applied bias magnetic field, with associated magnetization hysteresis. We tasked a series of agents, working within PhysKit, with reconstructing the per-island magnetization reversal of a pinwheel ASI \cite{gliga_emergent_2017} consisting of 2,178 nanomagnets from single-component confocal NV magnetic imaging data (Fig. \ref{fig:case-asi}b).  This problem was challenging for two reasons. First, the problem is ill-posed as the NV/ASI standoff of $\sim$700~nm and pixel pitch of 500 nm suppress the fine magnetic field features distinguishing individual island states at the nanoscale by a factor of $\sim$$10^4$--$10^5$.  

Second, the reconstruction involves roughly a dozen imperfectly known instrumental and physical parameters and several interdependent subproblems that have to be solved in order to progress to the final solution. Progress on each subproblem depended on the others; for example, errors in the standoff or NV confocal microscope point spread function (PSF) can be partially absorbed by the fit or bias magnetic field gradient subtraction. Arriving at the final optimization algorithm (Fig.~\ref{fig:case-asi}c) ultimately required a close, collaborative effort between the human users and a series of agents over approximately a dozen separate campaigns and interactive sessions spanning two months, with extensive discussion between agents and users. To our knowledge, the resulting solution provides the first reconstruction of the Ising microstate of an ASI array from diffraction-limited magnetic imaging data; a thorough treatment of these results will be included alongside experimental details in an upcoming manuscript \cite{everaert_imaging_2026}.

The forward model used in ASI magnetization reconstruction was established piece by piece during successive campaigns: fitting experimental NV data to determine the NV/ASI standoff, NV confocal microscope PSF, and the sensing NV azimuthal angle in the lab frame; including per-frame lattice rotation; accounting for field gradients caused by the bias magnet; and modeling the Stoner--Wohlfarth canting of the nanomagnet moments \cite{stoner_mechanism_1948}, which agents validated independently with MuMax3 \cite{vansteenkiste_design_2014} micromagnetic simulations  they designed and ran themselves. Removing the far-field components of the measured and predicted magnetic fields before fitting ensured that the fit was driven by the local field contribution of each nanomagnet island and also helped suppress field gradients induced by the bias magnet and other effects. Early campaigns tested direct CNN reconstruction of the per-frame magnetization maps: the networks reconstructed synthetic data well, but their $\sim$ $10^5$ unconstrained output values (per-pixel vector magnetization across all frames at $53\times53$ resolution) could not be justified for an ill-posed inversion. 

The successful ``joint fit'' approach takes advantage of many physical constraints, including the known geometry and the monotonic behavior of the islands during magnetization hysteresis (each island can switch its magnetization no more than once as the bias magnetic field is increased). This approach improves reconstruction accuracy while eliminating unphysical solutions.\footnote{The ``joint fit'' approach was previously proposed by our team more than a year prior to the collaborative agentic campaign, but could not be implemented due to computational limitations at that time.  It was later suggested independently by agents working on this project.} The final joint fit optimization algorithm employs the forward model and registration procedure developed by previous campaigns and uses warm starts and multi-seed fits to maximize the coefficient of determination $R^2$ between the predicted and measured magnetic field data, achieving a final $R^2$ of 0.9 and reducing the rate of misattributed Ising states to under 2\%. The estimated per-island switching uncertainty from the experimental data is 0.5~mT. 

We include this case study as an example of a complex, multi-campaign project for which, unlike the other case studies, we cannot show a single metric trajectory, because the constituent campaigns targeted different subproblems and used different metrics. Figure~\ref{fig:case-asi}c instead compares the reconstruction at the start and end of the project, showing our first direct fit attempt (``baseline fit'') and the final ``joint fit'' for three selected frames from the 27-frame hysteresis series.

\subsection{Lessons from Case Studies}
\label{sec:case-lessons}
The above case studies demonstrate that collaborative agentic research is capable of making meaningful contributions to multiple projects in our laboratory; and we expect that the principles and techniques discussed here will generalize to many areas of experimental physics. In particular, the results from the presented case studies directly surpass our previously published work in a variety of ways, including performance, fit model agreement with experimental data, and time required to reach an optimal solution. Agents frequently found creative solutions to a wide range of problems, handled debugging in real time, and designed well-targeted follow-up experiments (both synthetic and with hardware) to test hypotheses. With carefully designed agentic workflows and scaffolding, we could substantially extend the complexity and difficulty of problems that LLMs can handle beyond what fits in a single context window. Across all closed-loop campaigns, a critical element, unsurprisingly, is defining the right metric for the problem; a well-chosen metric functions as an effective filter for most errors or hallucinations that the agent makes. For research problems in this field, an effective metric must often include several connected objectives paired with a set of carefully chosen constraints, in order to prevent unintentional reward hacking. For especially complex problems, multiple campaigns with distinct objectives and workflows may be required, as seen in our ASI case study (Sec.~\ref{sec:case-asi}).  The lessons learned from these case studies have been incorporated into the PhysKit research framework~\cite{reed_physkit_2026}.

\section{Discussion}
\label{sec:discussion}

We have presented a principles-based approach to collaborative agentic research in experimental physics, together with a framework (PhysKit) that codifies these principles into agent-driven loop-based workflows using various skills tailored to experimental work. In six case studies, we demonstrate the applicability of agentic research to diverse tasks, including interpreting experimental data, optimizing quantum control protocols, designing hardware devices, and controlling an active quantum sensing experiment. We now discuss three themes: the broader impact of agentic workflows on experimental physics, their implications for the role of human researchers, and a future outlook.

\textbf{Impact on experimental physics.} Agentic workflows are fundamentally transforming experimental physics research across diverse tasks and problems. Previously, establishing whether a new research direction was viable required days or weeks of coding and testing on laboratory instrumentation, inhibiting rapid exploration of promising ideas. The low marginal cost of agentic research assistants enables efficient pursuit of lines of inquiry that would previously have been set aside, allowing human researchers to explore a wider range of opportunities and ideas, home in on the best options, and pursue them in greater depth. In our laboratory, implementing agentic workflows has resulted in measurably improved device designs, more rigorous data analyses, as well as a wealth of new theoretical and experimental directions that are benefiting our work. However, effectively using this capacity requires caution: the ability to quickly test the feasibility of nearly any idea with only a few prompts means that it is possible to waste time and resources on underdeveloped or poorly conceived ideas if the researcher is unfamiliar with the relevant material or fails to dedicate sufficient attention to the project.  

The long-term impact of agentic research on experimental physics remains to be determined. One possibility is that human researchers will be tempted to focus on rapidly producing incremental results, for which agents are well suited, rather than investing effort in fewer, more groundbreaking ideas and challenging experiments. Indeed, early AI adopters appear to be successful in increasing their number of publications, but rarely open new research directions \cite{haoArtificialIntelligenceTools2026}, although these early assessments may not carry over to later adopters or to more deliberate practice as the use of AI matures. In the long term, the ability of agents to accelerate useful, albeit incremental, experimental progress and identify subtle patterns in existing data may lay the groundwork that inspires innovative discovery. 

\textbf{Evolving role of the human researcher.} Collaborative agentic research effectively shifts the role of human researchers from doing labor-intensive software and experimental optimization tasks toward generating ideas, devising research strategies, orchestrating agents, assessing data, and verifying results. For the six case studies described in Sec.~\ref{sec:cases}, agentic research campaigns were initially based on human researcher specification and often required human redirection or iteration, such as questioning a dubious assumption, recommending a search direction, or trying different models, workflows, or research strategies if the agent failed to progress. Additionally, some campaigns uncovered findings that call into question the specified premise or metric, requiring further investigation and judgment from the human researcher. In all our case studies, we found that detailed verification and assessment of agentic results by human researchers is an essential task, generally replacing technical implementation as the limiting step of a research project. Even when work is well documented, verifying that the conclusions of agentic research are suitable for peer-reviewed publication demands considerable time and expertise from human researchers. As both direction of agents and verification of their work products require  knowledge and intuition, understanding how to train students to use agentic tools and make such decisions effectively will become increasingly important. 

A prominent strand of recent work centers on fully autonomous ``AI scientists'' designed to operate with minimal human direction through the entire research process, from idea generation through paper drafting \cite{lu_ai_2024,tang_airesearcher_2025,tang_fars_2026}. There are reasons to be skeptical that this approach is currently appropriate for experimental physics research. First, we find that agents struggle to exhibit scientific ``taste'': evaluating which questions are worth exploring in the first place, when to give up on an idea or explore it further, or how to shape results into a compelling narrative for a high-quality publication; our experience is consistent with emerging literature that identifies these areas as particular weaknesses \cite{bisht_agentic_2026,kirgis_can_2026}. Second, human verification is important for maintaining an accurate scientific record, particularly in empirical sciences where theories can only be tested statistically, and accounting for systematic bias and uncertainty can require significant judgment and expertise. Third, while AI can implement and optimize experimental procedures, as demonstrated in the NMR case study (Sec. \ref{sec:case-nmr}), automating the construction of new experiments is currently out of reach. We therefore prefer a workflow where collaborative agents provide technical and analytic support that enhances our ability to make expert-level judgments, rather than handing such judgments to AI. Our ASI reconstruction case study (Sec.~\ref{sec:case-asi}) is an example of what this process can look like: it took a dozen campaigns and many discussions between human researchers and agents to refine the final code, with many key insights arising through human--agent interaction in response to campaign results. 

\textbf{Future outlook.} We expect collaborative agentic research to affect most areas of experimental physics (fundamental and applied), reshaping workflows and the skills required of human researchers. As its use spreads, establishing scientific norms for AI behavior becomes central, both for human researchers using agentic tools and for developers designing scientific harnesses and post-training refinements.  The principles of Sec.~\ref{sec:rules} restate norms that predate computing, while the files, hooks, and loops that enforce them track the capabilities and failure modes of current agentic models.  We expect these techniques to continue to evolve, e.g., via exploration of file customization for harnesses optimized for the diverse activities of experimental physics. 

As of this writing, clear standards for documentation, reporting, and verification of AI work in physics (and the natural sciences more broadly) do not exist. Once established, such standards could be enforced more systematically, for example, through journal review processes, stricter data and agentic model availability requirements, and robust community standards \cite{Ceriotti_Editorial_2026}. Additionally, validated benchmarks are needed to quantify agentic performance on realistic research problems in experimental physics, so that different training protocols or frameworks can be compared. For example, experimental physicists in precision measurement, rare-event searches, and particle physics regularly blind their analyses to protect results from human bias \cite{klein_blind_2005,maccoun_blind_2015,ayres_data_2021,lzcollaboration_dark_2025}; as AI agents become increasingly common in experimental physics research, we should extend the same care to ensure they strengthen rather than erode the reliability of the scientific record. 

\section{Methods}

All results were generated using commercially available LLMs and agentic harnesses (Claude Code and Codex) and supported by our custom framework, PhysKit. Case study A (current density reconstruction) used Claude Opus 4.8 and PhysKit 0.6.0; case study B (improved photodynamic model for $V_B^-$ spin defects in hBN) used GPT 6 Astra and PhysKit 0.12.1; case study C (microwave device design) used Claude Opus 5 and PhysKit 0.6.0; case study D (pulse sequence design) used Claude Fable 5 and PhysKit 0.6.0, with additional analytic reasoning provided by ChatGPT 5.6 Sol at Pro thinking effort; case study E (NMR experiment commissioning) used Claude Fable 5.1, Opus 5.5, and GPT 6 Astra with PhysKit 0.12.1; case study F (artificial spin-ice magnetization reconstruction) comprised several distinct campaigns and sessions and used Claude Sonnet 4.5, Sonnet 5, Opus 4.8, Opus 5, and Fable 5, with PhysKit 0.3.0, 0.4.0, and 0.6.0. PhysKit was implemented as a custom plugin for both Claude Code and Codex, including custom skills, a system prompt, and campaign templates (available at Ref.~\cite{reed_physkit_2026}). AI agents were used to assist with data analysis and plotting in the case studies. Simulation tasks in the case studies used the physics simulation packages MuMax3 and Ansys HFSS. Experimental case studies used data taken on custom-built NV and $V_B^-$ quantum sensing microscopes; see appendix for additional details.

\begin{acknowledgments}
We thank George Witt, Gavishta Liyanage, Maximilian Shen, Chinmay Bharathulwar, William Taylor, and other Walsworth lab members for useful discussions on deploying agentic AI in research. This work was supported in part by the National Science Foundation Graduate Research Fellowship Program under Grant No. DGE 2236417; the Laboratory for Physical Sciences Jumping Electron Quantum Fellowship Program under Award No. H9823022C0029; the U.S. Army Research Laboratory under Contract No. W911NF2420143; the U.S. Air Force Office of Scientific Research under Grant No. FA9550-22-1-0312 as part of the MURI project Comprehensive Minimally/non-invasive Multifaceted Assessment of Nano-/microelectronic Devices (CoMMAND); the Gordon \& Betty Moore Foundation under Grant No. 7797.01; the Maryland Procurement Office under Award No. H9823019C0220; and the University of Maryland Quantum Technology Center.

\end{acknowledgments}

\bibliography{main}
\end{document}